\documentclass[aps,showpacs,twocolumn,pre,10pt]{revtex4}
\usepackage{bm}
\usepackage{amsfonts}
\usepackage{amssymb}
\usepackage{graphicx}
\usepackage{amsmath}
\usepackage{indentfirst}

\begin{document}
\title{Annular billiard dynamics in a circularly polarized strong laser field}

\author{A. Kamor$^1$, F. Mauger$^{2}$, C. Chandre$^2$, and T. Uzer$^1$}

\affiliation{$^1$ School of Physics, Georgia Institute of Technology, Atlanta, GA 30332-0430, USA\\
$^2$ Centre de Physique Th\'eorique, CNRS -- Aix-Marseille Univ, Campus de Luminy, case 907, 13288 Marseille cedex 09, France}

\begin{abstract}
We analyze the dynamics of a valence electron of the buckminsterfullerene molecule (${\rm C}_{60}$) subjected to a circularly polarized laser field by modeling it with the motion of a classical particle in an annular billiard. We show that the phase space of the billiard model gives rise to three distinct trajectories: ``Whispering gallery orbits'',  which only hit the outer billiard wall, ``daisy orbits'' which hit both billiard walls (while rotating solely clockwise or counterclockwise for all time), and orbits which only visit the downfield part of the billiard, as measured relative to the laser term.  These trajectories, in general, maintain their distinct features, even as intensity is increased from $10^{10}$ to $10^{14} \ {\rm W}\cdot{\rm cm}^{-2}$.  We attribute this robust separation of phase space to the existence of twistless tori.
\end{abstract}


\maketitle
\section{Introduction} \label{sec:intro}

The electrical and chemical properties of fullerenes, namely ``buckyballs'' and nanotubes, remain the focus of thorough investigations~\cite{Hert05}. The Buckminsterfullerene molecule, ${\rm C}_{60}$, is a prototypical nanocluster because of its stability and nearly spherical shape. In particular, it is an ideal cage in which to trap so-called endohedral atoms, resulting in molecular systems with peculiar properties~\cite{Forr01}, i.e., enhanced stability with respect to temperature. In recent years, there has been a significant interest in subjecting ${\rm C}_{60}$ to extreme conditions to probe its electronic and structural stability properties.  A new class of experiments on ${\rm C}_{60}$ driven by strong laser pulses show that its ionization and fragmentation properties are very sensitive to the laser intensity and polarization~\cite{Hert09}. In particular, the yields show remarkable changes with the ellipticity of the laser field. 

Motivated by these findings, we consider the motion of an electron inside the valence shell of ${\rm C}_{60}$. The goal is to understand the electronic dynamics prior to photoionization.  In strong linearly polarized laser fields, the ionized electron can return to the remaining ion by recolliding with the cage~\cite{Cork93,Bhar04} when the laser field changes sign. This collision can lead to additional ionization or even fragmentation~\cite{Hert09} of the molecule.  

In particular, we investigate the classical dynamics of a one-electron model in the shell of fullerene ${\rm C}_{60}$ subjected to a strong circularly polarized laser field.  In a circularly polarized field, the dynamics is best visualized in a frame co-rotating with the laser field where a conserved quantity emerges, the Jacobi constant~\cite{Hill78}.  All results presented in this paper are in the rotating frame in which the circularly polarized laser becomes a static field with definite upfield and downfield directions. We restrict the dynamics to a two-dimensional configuration space (the plane of polarization) for the valence electron. The valence electron feels an averaged potential, which, as we later show (Sec.~\ref{sec:Ham_model}), is very close to a spherical square well potential, where the electron bounces between the walls like a particle in an annular billiard. We choose a billiard model for its simplicity both analytically and numerically and because it serves as a faithful representation of the full model potential while not allowing ionization to occur. 

Annular billiards occur in the literature in at least two contexts: Fermi acceleration and the study of quantum chaos by comparison with classical and quantum mechanical computations~\cite{Doro95,Kohl98,Robi99,Demb00,Hent02,Goue01,Egyd06_1,Egyd06_2}. Chaotic dynamics arises either from pulsating boundaries or from an off-center inner wall. In our treatment the two walls are fixed and concentric. The main distinction here from other works on annular billiards is that in the rotating frame the electron moves along curved paths between successive wall collisions. The introduction of a Coriolis term into the Hamiltonian upon the transformation to the rotating frame is, of course, akin to introducing an effective uniform magnetic field (and another frequency, the Larmor frequency) in which an electron moves on a curved path~\cite{Robn85,Mepl93,Berg96,Silv00,Aich10}. The laser wavelength is taken as 780nm (corresponding to a frequency of 0.0584~a.u.) and its intensity is varied from zero to $10^{14} \ {\rm W}\cdot{\rm cm}^{-2}$, which are values consistent with what is routinely performed in experiments on ${\rm C}_{60}$~\cite{Hert05}. 

Our principal finding is that trajectories of the electron fall into one of three possible types which originate from specific parts of phase space. We identify various phase space structures which keep these trajectory types distinct from one another.  Our classification is as follows: First there are ``whispering gallery orbits'' (WGO).  These trajectories hit only the outer wall and their direction of travel, either clockwise or counterclockwise, is determined by which of the two regions in phase space where they originate.  The second type, which we call ``daisies'' from their typical shape, hits both walls and can rotate either clockwise or counterclockwise.  Both daisies and WGOs are called positive or negative based upon their direction of travel, counterclockwise or clockwise, respectively.  The third type is mainly influenced by a very simple elliptic periodic orbit bouncing between the two walls in the downfield direction.  The shape of these trajectories on the Poincar\'{e} section resembles a popular snack food, the Pringles curved potato chip~\cite{pringle}, thus their designation as ``pringle orbits.'' 

We show that all the trajectories maintain their distinct characteristics with changing intensity because of a special class of invariant tori which do not fulfill the usual twist hypothesis required by the Kolmogorov-Arnold-Moser (KAM) theorem. These tori are usually denoted as ``twistless'' tori or shearless curves (of the Poincar\'e map) or ``meandering'' curves when they are associated with separatrix reconnection~\cite{Morr09, Negr96, Morr00}. Numerical studies show that these tori are very robust against perturbation, and are natural candidates for the partitioning of phase space into regions where well-defined, qualitatively distinct trajectories can be found.	

The plan of the article is as follows: In Sec.~\ref{sec:dynRules}, we introduce the model and dynamical rules of the billiard and its corresponding Hamiltonian and reflection guidelines. In Sec.~\ref{sec:analysis}, we give a qualitative analysis of different electron trajectories, and then relate them to the organization of phase space. Next, we analyze the dynamics by using Poincar\'e section and frequency map analysis~\cite{Lask93} in order to gain deeper insight into the properties and organization of phase space when the laser intensity is increased, and in particular concerning the role of twistless tori in the partitioning of phase space into three principal regions.   


\section{Dynamical Rules}\label{sec:dynRules}

\subsection{Hamiltonian Model} \label{sec:Ham_model}
\begin{figure}
   \centering
   \includegraphics[width = \linewidth]{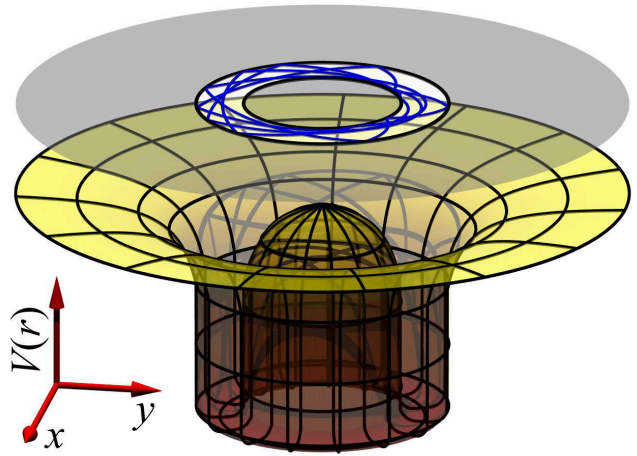}
   \caption{\label{fig:potential_diagram} Potential {\it V}, felt by a valence electron in fullerene as given in Ref.~\cite{Shch06_2}.  The overhead plane corresponds to the accessible billiard region (white space) with a sample trajectory (blue curve).}
\end{figure}

A typical electron inside the shell feels the influence of the electrostatic potential created both by the positive charges from the nucleus and the electronic density, combined with the influence of the laser field. The effective single-particle potential is computed using density functional theory from a jellium approximation for the positive charge background~\cite{Pusk93}. It contains steep walls in the potential around the radius of fullerene ($r_0=6.69$ a.u.)~\cite{Pusk93,Baue01,Rugg08,Shch06_2} with a certain thickness~\cite{Balt10}. The Hamiltonian expressed in atomic units (a.u.) and in the dipole approximation, reads
\begin{equation} \label{eq:fullModel_Ham}
   \mathcal{H} \left({\bf x}, {\bf p}, t \right) = \frac{\left|{\bf
p}\right|^{2}}{2} + V\left(\left|{\bf x}\right|\right) + {\bf x}\cdot{\bf
E}\left(t\right)
\end{equation}
where ${\bf x}=\left(x,y\right)$ is the position of the electron in the polarization plane, ${\bf p}=\left(p_{x},p_{y}\right)$ its canonically conjugate momenta and $\left|\cdot\right|$ denotes the Euclidean norm. The circularly polarized laser field is given by ${\bf E}(t) = E_{0}({\bf e}_{x}\sin\omega t+{\bf e}_{y}\cos\omega t)$, where $E_{0}$ is the electric field amplitude, $\omega$ is the laser frequency, kept fixed at 0.0584 a.u., and ${\bf e}_x$ and ${\bf e}_y$ are unit vectors along the $x$ and $y$ axes, respectively.  The laser intensity is the time averaged Poynting vector of our laser field and is related to $E_0$ by the relationship $I = \alpha E_0^2$, where $\alpha= 7.044\times 10^{16}$ when laser intensity is measured in ${\rm W}\cdot{\rm cm}^{-2}$.  Figure~\ref{fig:potential_diagram} shows the potential~$V(r)$, where $r=\vert {\bf x}\vert$ as given in Ref.~\cite{Shch06_2}. We note that the potential is very stiff at the boundaries of the shell. This is a common feature of various models for ${\rm C}_{60}$~\cite{Pusk93,Baue01,Rugg08}. This property holds for ions ${\rm C}_{60}^{q+}$~\cite{Baue01} also.  An approximate potential consisting of a spherical square well potential, where the potential is equal to $-V_0$ for $r\in]r_0-\delta,r_0+\delta[$ and zero elsewhere, has been proposed in Refs.~\cite{Xu96,Tosa94}. This model has succeeded in explaining the oscillations in the photoionization cross-section of ${\rm C}_{60}$~\cite{Xu96}.
We build a billiard model along these lines, where the steep walls of the potential are replaced with infinite walls and the dynamics in the annular region between the two walls is given solely by the interaction of the electron with the electric field, later referred to as the laser-driven dynamics:  
\begin{equation} \label{eq:billiardStatic_Ham}
    \mathcal{H} \left({\bf x}, {\bf p} \right) = \frac{\left|{\bf p}\right|^{2}}{2} + {\bf x}\cdot{\bf E}\left(t\right),
\end{equation}
and reflection rules are applied whenever the trajectory reaches $r=r_{\rm in}$ or $r_{\rm out}$, which are 5.14 a.u. and 8.24 a.u., respectively in our computations.

First, we perform a canonical change of variables into a rotating frame (with the laser field). The new coordinates $\left(\overline{x},\overline{y},\overline{p}_{x},\overline{p}_{y}\right)$ are given by
$$
   \left(\begin{array}{c} \overline{x} \\ \overline{y} \end{array}\right) = {\bf \Omega}\left(t\right)\left(\begin{array}{c} x \\ y \end{array}\right), \, \mbox{
and } 
   \left(\begin{array}{c} \overline{p}_{x} \\ \overline{p}_{y} \end{array}\right)={\bf\Omega}\left(t\right)\left(\begin{array}{c} p_{x} \\ p_{y} \end{array}\right), 
$$
where
$$
   {\bf \Omega} \left( t\right) = \left(\begin{array}{cc} \sin \omega t & \cos \omega t \\ \cos \omega t & - \sin \omega t \end{array}\right).
$$
In the new set of variables, the Hamiltonian becomes time-independent and reads
\begin{equation} \label{eq:Ham}
   \mathcal{K} \left(x, y, p_{x}, p_{y}\right)  = \frac{p_x^2}{2} + \frac{p_y^2}{2} - \omega \left(x p_y - y p_x\right) + E_0 x, 
\end{equation}
where we have dropped the bars for simplicity. The resulting Hamiltonian has two degrees of freedom and the value of the Hamiltonian is the Jacobi
constant of celestial mechanics~\cite{Hill78}.

\subsection{Topology of phase space}\label{sec:topology}

\begin{figure}
   \includegraphics[width = \linewidth]{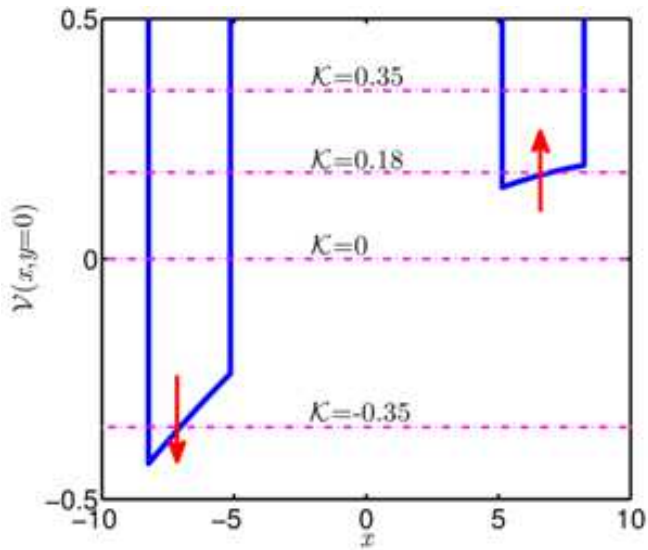}
   \caption{\label{fig:zvs}  Section ($y=0$) of the zero-velocity surface in the accessible part of the billiard for $I=10^{14} \ {\rm W}\cdot{\rm cm}^{-2}$ and $\omega=0.0584$ a.u. The red arrows show the deformation of the zero velocity surface as intensity is increased.  The dashed horizontal lines are the Jacobi values used in Fig.~\ref{fig:accessibleRegions} and Fig.~\ref{fig:varyK}.}
\end{figure}

\begin{figure}
  \centering
  \includegraphics[width = .49 \linewidth]{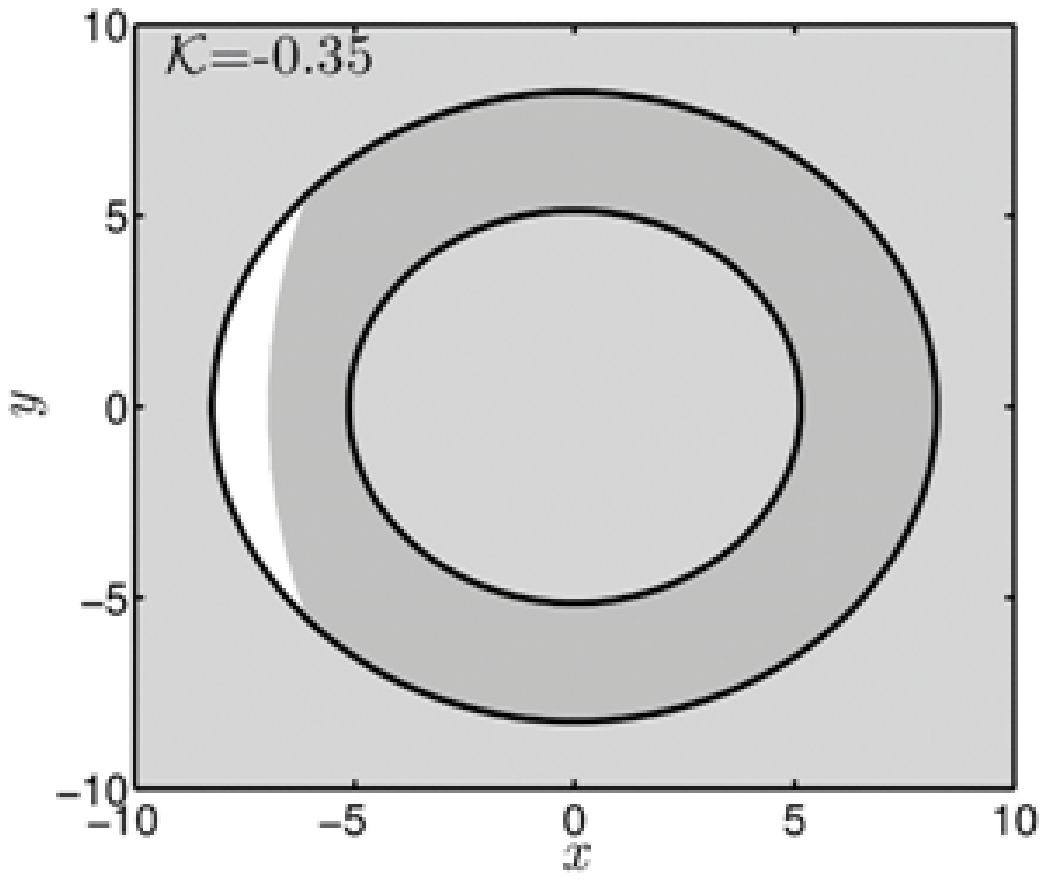}
  \includegraphics[width = .49 \linewidth]{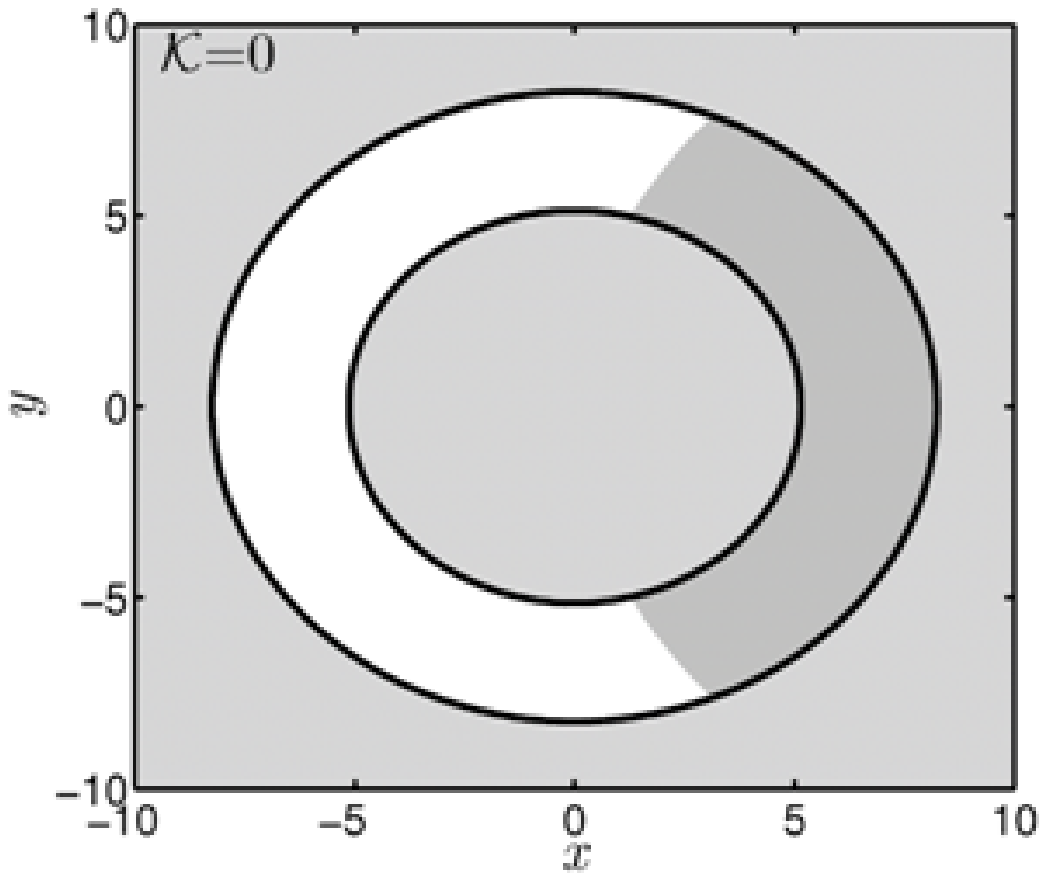}
  \includegraphics[width = .49 \linewidth]{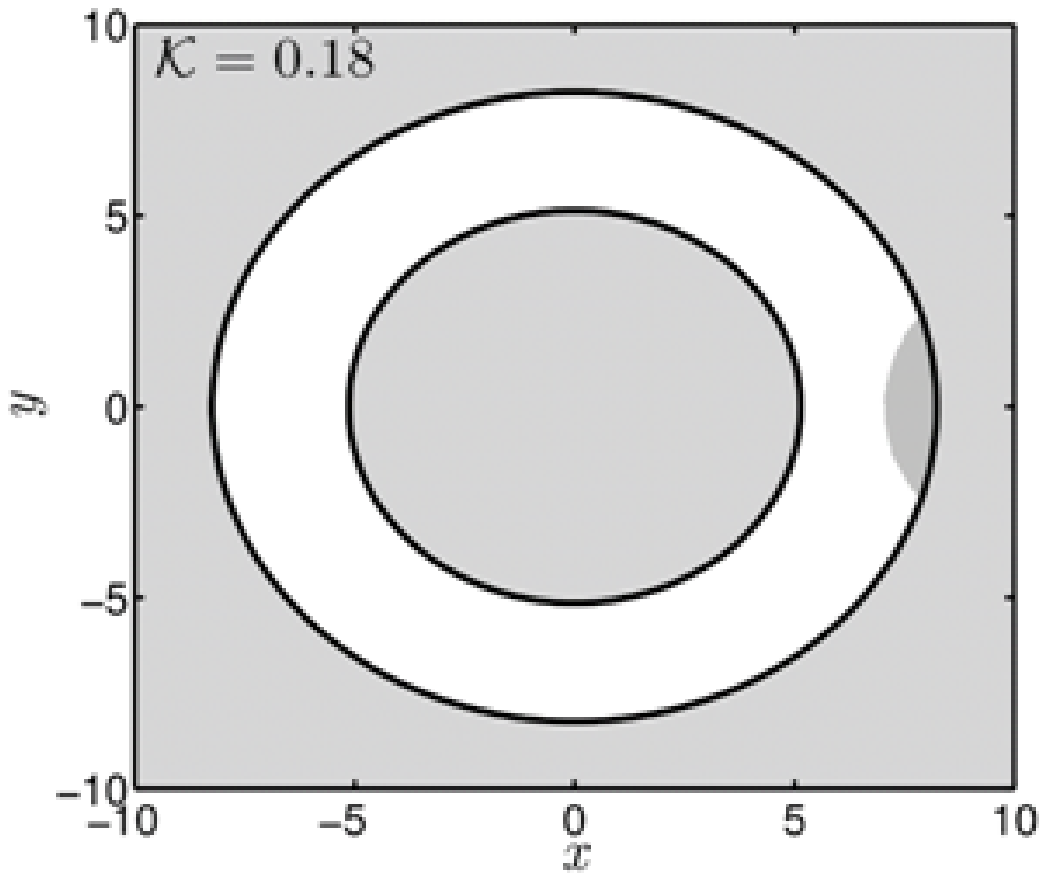}   
  \includegraphics[width = .49 \linewidth]{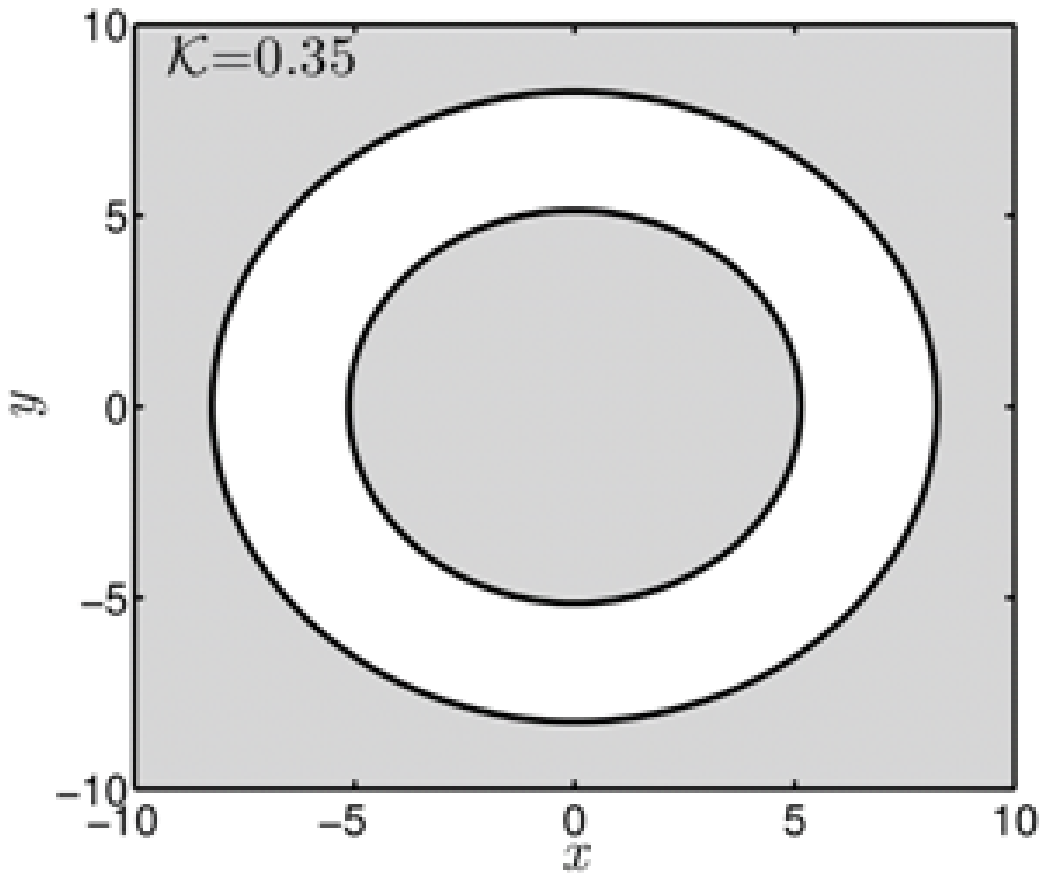}
  \caption{Accessible regions (white) of the billiard for different Jacobi values at $I=10^{14} \ {\rm W}\cdot{\rm cm}^{-2}$ and $\omega=0.0584$ a.u. Jacobi values correspond to the dashed lines in Fig.~\ref{fig:zvs}.}
\label{fig:accessibleRegions}
\end{figure}

The accessible part of phase space changes depending on the field frequency, amplitude, and the value of the Jacobi constant. A revealing way to visualize the accessible part in position space is to compute the zero velocity surface~\cite{Hill78}. By applying Hamilton's equations to Eq.~(\ref{eq:Ham}) we arrive at

\begin{eqnarray*}
   \dot x &= p_x + \omega y \label{eq:xDot}, \\
   \dot y &= p_y - \omega x \label{eq:yDot},
\end{eqnarray*}
so that the Jacobi constant becomes
\begin{equation}
   \mathcal{K}\left(x,y,\dot{x},\dot{y}\right) = \frac{\dot{x}^2}{2} + \frac{\dot{y}^2}{2} - \frac{1}{2}\omega^2\left(x^2 + y^2\right) + E_0 x.
\end{equation}
Setting $\dot{x}=\dot{y}=0$ gives the zero-velocity surface
$$
   \mathcal{V}\left(x,y\right) = - \frac{1}{2}\omega^2\left(x^2 + y^2\right) + E_0 x,
$$
which charts the lower limit of the accessible parts of the billiard, as $\mathcal{K}$ is varied. A cross section of the zero-velocity surface is shown in Fig.~\ref{fig:zvs} for $y=0$.  
Depending on the value of ${\cal K}$, three possibilities arise: If ${\cal K}$ is smaller than $-\omega^2 r_{\rm out}^2/2-E_0 r_{\rm out}$, then there is no accessible part to the dynamics. If ${\cal K}$ is between $-\omega^2 r_{\rm out}^2/2-E_0 r_{\rm out}$ and $-\omega^2 r_{\rm in}^2/2+E_0 r_{\rm in}$ then only a portion of the annular region is accessible.  In this range of values, several truncations of the annulus are possible; in particular, we distinguish two types: one which is homotopic to an annulus, and one which is only a portion of an annulus (see Fig.~\ref{fig:accessibleRegions}).  For ${\cal K}$ larger than $-\omega^2 r_{\rm in}^2/2+E_0 r_{\rm in}$, the entire annulus is accessible to the dynamics. As the laser intensity is increased, the difference between the right and left sides of the well is amplified. In this paper we mainly consider Jacobi constants larger than $-\omega^2 r_{\rm in}^2/2+E_0 r_{\rm in}$ such that the full annular region of the billiard is accessible to the dynamics. For $I=10^{14} \ {\rm W}\cdot{\rm cm}^{-2}$ and $\omega=0.0584$ a.u., this critical value of ${\cal K}$ is approximately equal to 0.15.  
 
\subsection{Dynamical rules}\label{sec:sub_dynRules}

The dynamics is computed in a piecewise fashion because of the walls. It is composed of segments of laser-driven dynamics, as given by
Hamiltonian~(\ref{eq:Ham}), until the particle reaches one of the walls. At this instant, the reflection rule is applied which mimics an elastic scattering at the limit of an infinitely stiff potential.

Concerning the laser-driven dynamics, the equations of motions associated with Hamiltonian~(\ref{eq:Ham}) are given by
\begin{subequations} \label{eq:laserDrivenDyn}
\begin{eqnarray}
x\left(t\right) &=& \frac{E_0}{\omega^2} + \left(\left(x_0-\frac{E_0}{\omega^2}\right)+p_{x,0}t\right)\cos{\omega t} \nonumber \\ 
                &&+\left( \left( p_{y,0}-\frac{E_0}{\omega}\right)t + y_0\right)\sin{\omega t}, \\  
y\left(t\right) &=& \left(y_0 + \left(p_{y,0}-\frac{E_0}{\omega}\right)t\right)\cos{\omega t} \nonumber \\
                &&- \left(\left(x_0 - \frac{E_0}{\omega^2}\right)+p_{x,0}t\right)\sin{\omega t}, \\
p_x\left(t\right) &=& \left(p_{y,0}-\frac{E_0}{\omega}\right)\sin{\omega t} + p_{x,0} \cos{\omega t}, \\
p_y\left(t\right) &=& \left(p_{y,0}-\frac{E_0}{\omega}\right)\cos{\omega t} -p_{x,0} \sin{\omega t} + \frac{E_0}{\omega}, 
\end{eqnarray}
\end{subequations}
where $x_0$, $y_0$, $p_{x,0}$, and $p_{y,0}$ are the initial conditions (at time $t=0$).

At time $t=t_{R}$, the electron reaches either one of the two walls and we apply the reflection condition before a next phase of laser-driven dynamics. The reflection conditions are more easily expressed in polar coordinates as it corresponds to the radial momentum changing sign while the other coordinates are unchanged (see Fig.~\ref{fig:ReboundLocalFrame}). By definition of the billiard, the rebound takes place at time $t=t_{R}$ satisfying
$$
   \sqrt{x^{2}\left(t_{R}\right) + y^{2}\left(t_{R}\right)} = r_{i},
$$
where $i$ denotes the wall index, i.e., $i\in\left\{{\rm in}, \ {\rm out}\right\}$. Before the rebound (i.e.,\ at $t=t_R^-$), we compute the radial
momentum $p_r$ from the value of $(x,y,p_x,p_y)$ by $p_r=(x p_x+y p_y)/r$. Then, the rebound condition is given by a change of sign of the radial momentum
$$
p_{r}\left(t_{R}^{+}\right)=-p_{r}\left(t_{R}^{-}\right).
$$
As a consequence, the values of the momenta after the rebound, denoted $p_x^+$ and $p_y^+$, are given by 
\begin{subequations} \label{eq:ReboundCondition}
  \begin{eqnarray}
      p_{x}^+ & = & -\frac{(x^2-y^2)p_x^-+2xyp_y^-}{r^2}, \\
      p_{y}^+ & = & \frac{-2xyp_x^-+(x^2-y^2)p_y^-}{r^2}.
  \end{eqnarray}
\end{subequations}
We note that Hamiltonian~(\ref{eq:Ham}) is left unchanged by the reflection rule, that is~$\mathcal{K}\left(t_{R}^{+}\right)=\mathcal{K}\left(t_{R}^{-}\right)$, which is easily seen from the expression of the kinetic energy in polar coordinates which is equal to~$p_{r}^2/2+p_{\theta}^{2}/\left(2r_{i}^{2}\right)$.


\subsection{Linearization of the flow}

In this section, we consider the linear effect of the rebound condition on neighboring trajectories. The motivation for doing so is twofold: First the tangent rebound condition can be used to compute the tangent flow of trajectories to deduce the linear stability of periodic orbits. Second, as the billiard model corresponds to the limit of infinitely stiff potential for a Hamiltonian system, the billiard model should preserve the Hamiltonian structure. We have already checked that the rebound condition preserves the Hamiltonian, thus the last prescription is that the symplectic two-form is preserved, or equivalently the tangent rebound matrix is symplectic. 

\begin{figure}
   \includegraphics[width = \linewidth]{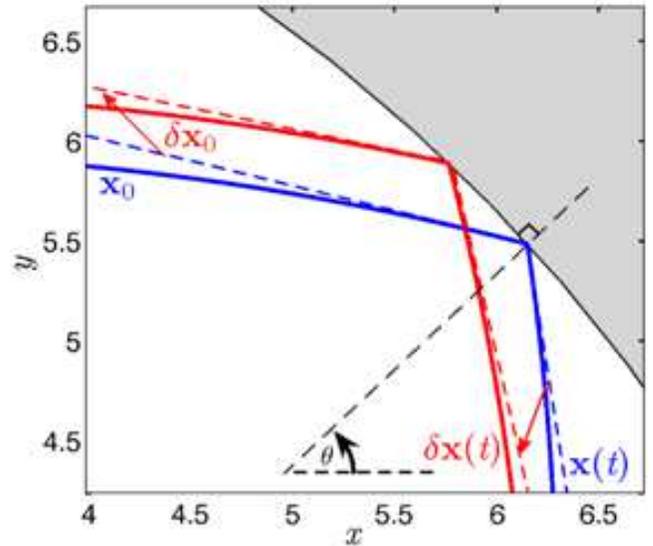}
   \caption{\label{fig:ReboundLocalFrame} 
   Schematic representation for computation of the rebound condition.  The solid blue and red lines are real trajectories, differing only by a small perturbation in the initial conditions.  The dashed red and blue lines are representations of the linearized dynamics. The angle $\theta$ is the angle of rotation used in Hamiltonian~(\ref{eq:RotatedHamilt}).}
\end{figure}

The tangent rebound matrix can be seen as an extension of the tangent flow~\cite{chaosbook} to include the impact of the rebound on neighboring trajectories (to first order). For that, we consider a first trajectory with initial conditions $x_{0}, y_{0}, p_{x,0}, p_{y,0}$ at $t=0$ in the neighborhood of one of the two walls, i.e., $x_{0}^{2}+y_{0}^{2}\approx r_{i}^{2}$, where $i\in\left\{{\rm in},\ {\rm out}\right\}$, and such that the rebound time $t_{R}\ll 1$. Then, we look at the impact of a perturbation of these initial conditions to $x_{0}+dx_{0}, y_{0}+dy_{0}, p_{x,0}+dp_{x,0}, p_{y,0}+dp_{y,0}$ immediately after the rebound. Because the rebound time is not the same for the two trajectories, we have to consider a small laser-driven propagation before and after the rebound to deduce the tangent rebound properties. The situation is schematically depicted in Fig.~\ref{fig:ReboundLocalFrame}.

The computation of the tangent rebound matrix is more easily seen in a rotated Cartesian set of coordinates $\tilde{x}, \tilde{y}, \tilde{p}_{x}, \tilde{p}_{y}$, for some angle~$\theta$ to be specified later. In the rotated frame, the corresponding Hamiltonian reads
\begin{eqnarray} \label{eq:RotatedHamilt}
   \mathcal{K}\left(x,y,p_{x},p_{y}\right) & = & \frac{p_{x}^{2}}{2} +
\frac{p_{y}^{2}}{2} - \omega\left(x p_{y} - y p_{x}\right) \nonumber \\
   & & + E_{0} x \cos \theta + E_{0} y \sin \theta, 
\end{eqnarray}
where we have dropped the tildes for simplicity. Since we are only interested in the linear properties of the rebound condition, throughout this section we consider all the equations linearized to the first order.

As explained in Sec.~\ref{sec:sub_dynRules}, the dynamics is computed piecewise with a first stage of laser-driven propagation until the electron reaches the wall, then the rebound condition and finally a new stage of laser-driven propagation. In the rotated frame and before the rebound, i.e., $t<t_{R}$, the laser-driven dynamics given by Hamiltonian~(\ref{eq:RotatedHamilt}) yields
\begin{subequations} \label{eq:TrajBeforeRebound}
   \begin{eqnarray}
      x \left( t \right)     & \approx & x_{0}   + \left( p_{x,0} + \omega y_{0}
\right) t , \label{eq:trajBeforeRebound_x}\\
      y \left( t \right)     & \approx & y_{0}   + \left( p_{y,0} - \omega x_{0}
\right) t , \\
      p_{x} \left( t \right) & \approx & p_{x,0} + \left( \omega p_{y,0} -
E_{0}\cos\theta \right) t , \\
      p_{y} \left( t \right) & \approx & p_{y,0} - \left( \omega p_{x,0} +
E_{0}\sin\theta \right) t,
   \end{eqnarray}
\end{subequations}
where we have neglected $O(t^2)$. At this stage, we select the rotation angle~$\theta$ such that the perpendicular direction to the wall at the rebound is aligned with the $x$-direction for the unperturbed trajectory. We keep this fixed frame for the perturbed trajectory. An alternative way (which provides the same solution) is to consider a perturbed rotated frame (obtained by a rotation by an angle $\theta+d\theta$) to impose the same constraint (the direction perpendicular to the wall is the $x$-axis at the rebound) on the perturbed trajectory. As a consequence of the chosen angle~$\theta$, the rebound condition solely depends on the $x$-direction, such that $x\left(t_{R}\right)=r_{i}$, or equivalently using Eq.~(\ref{eq:trajBeforeRebound_x})
\begin{equation} \label{eq:ReboundTime}
   t_{R} \approx \frac{r_{i} - x_{0}}{p_{x,0} + \omega y_{0}}.
\end{equation}
Besides, because of the circular shape of the billiard and the angle $\theta$, at the rebound the $y$-component vanishes ($y\left(t_{R}\right)=0$). However, for the perturbed trajectory, this condition does not apply. Since we consider the perturbed trajectory comparatively to the original one in the same rotated frame, it is easier for the purpose of the calculation to keep formally the $y$-components, knowing that it is actually equal to zero.

The next step for the trajectory dynamics is the rebound condition, at time $t=t_{R}$. Because of the orientation of the frame where the $x$-direction is aligned with the radial one, the rebound condition~(\ref{eq:ReboundCondition}) becomes~$\dot{x}\left(t_{R}^{+}\right)=-\dot{x}\left(t_{R}^{-}\right)$ and $\dot{y}\left(t_{R}^{+}\right)=\dot{y}\left(t_{R}^{-}\right)$ which implies to the momenta
\begin{equation} \label{eq:RotatedReboundCondition}
   p_{x}\left(t_{R}^{+}\right) = -p_{x}\left(t_{R}^{-}\right) - 2 \omega
y\left(t_{R}\right),
\end{equation}
and $p_{y}\left(t_{R}^{+}\right)=p_{y}\left(t_{R}^{-}\right)$, while the positions are left unchanged.

Finally, after the rebound, the trajectory experiences a new phase of laser driven dynamics. Combining the rebound condition~(\ref{eq:RotatedReboundCondition}) with a linearized propagation similar to Eq.~(\ref{eq:TrajBeforeRebound}), one can write the dynamics after the rebound as a function of the initial conditions (before the rebound), such that
\begin{subequations} \label{eq:AfterReboundDynamics}
   \begin{eqnarray}
      x \left( t \right)  & \approx &   x \left(t_{R}\right) + \left(p_{x}\left(t_{R}^{+}\right) + \omega y\left(t_{R}\right) \right) \left(t-t_{R}\right), \nonumber \\
                          & \approx &   x_{0} - \left(p_{x,0} + \omega y_{0}\right) \left( t - 2 t_{R} \right), \\
  y \left( t \right)      & \approx &   y_{0} + \left( p_{y,0} - \omega x_{0} \right) t, \\
  p_{x} \left( t \right)  & \approx & - \left(p_{x,0} + 2 \omega y_{0} \right) -2 \omega \left( p_{y,0} - \omega x_{0} \right) t_{R} \nonumber \\
                          &         & + \left( \omega p_{y,0} - E_{0} \cos\theta \right) \left( t -2 t_{R}\right), \\
  p_{y} \left( t \right)   & \approx &   p_{y,0} + 2 \omega \left( p_{x,0} + \omega y_{0} \right) \left( t - t_{R} \right) \nonumber \\
                           &         & - \left( \omega p_{x,0} + E_{0} \sin\theta \right) t,
   \end{eqnarray}
\end{subequations}
where the equations are linearized at the first order in time.  With the explicit formula for the dynamics after the rebound, as given by Eq.~(\ref{eq:AfterReboundDynamics}), it is straightforward to compute the impact of the perturbation on the initial conditions by replacing $x_{0}, y_{0}, p_{x,0}, p_{y,0}$ with $x_{0}+dx_{0}, y_{0}+dy_{0}, p_{x,0}+dp_{x,0}, p_{y,0}+dp_{y,0}$ respectively. Because of the change of initial conditions, the rebound time is modified to $t_{R}+dt_{R}$ as well. Using Eq.~(\ref{eq:ReboundTime}) for the perturbed trajectory we end up with
\begin{equation}\label{eq:PerturbedReboundTime}
   dt_{R} = - \frac{dx_{0}}{p_{x,0}+\omega y_{0}} - \frac{r_{i} - x_{0}}{\left(p_{x,0}+\omega y_{0}\right)^{2}} \left( dp_{x,0} + \omega dy_{0} \right).
\end{equation}
Finally, combining Eq.~(\ref{eq:AfterReboundDynamics}) with Eq.~(\ref{eq:PerturbedReboundTime}) it is possible to compute the perturbed dynamics after the rebound. We define the deviations $dx, dy, dp_{x}, dp_{y}$ of the perturbed trajectory after the rebound. For instance, considering the
$x$-coordinate, we obtain
\begin{eqnarray} 
   dx  & = & - dx_{0} - \frac{2 \left(r_{i}-x_{0}\right)}{p_{x,0}+\omega y_{0}}
\left( dp_{x,0} + \omega dy_{0} \right) \nonumber \\
       &         & - \left( dp_{x,0} + \omega dy_{0} \right) \left( t - 2 t_{R}
\right). \label{eq:x-dir_perturbation}
\end{eqnarray}
Since we are interested in the dynamics in the vicinity of the rebound, we consider the limits $x_{0}\rightarrow{r_{i}}$ and $t\rightarrow{t_{R}^{+}}$ (such that $t\rightarrow{0}$). As a consequence, we end up with $dx=-dx_{0}$ from Eq.~(\ref{eq:x-dir_perturbation}). A similar procedure can be applied to the other components and summarized in the linear equation
$$
\left(\begin{array}{c}  dx \\ dy \\ dp_{x} \\ dp_{y} \end{array}\right) =  J_{R} \left(\begin{array}{c}  dx_{0} \\ dy_{0} \\ dp_{x,0} \\ dp_{y,0} \end{array}\right),
$$
where $J_{R}$ is the tangent rebound matrix given by
\begin{equation} \label{eq:TangentReboundMatrix}
    J_{R} = \left( \begin{array}{cccc}
      -1 & 0 & 0 & 0 \\
       0 & 1 & 0 & 0 \\
       \left(4 \omega p_{y} - 2 E_{0}\cos\theta - 2 \omega^{2} r_{i}\right)p_{x}^{-1} & -2 \omega & -1 & 0 \\
       2 \omega & 0 & 0 & 1
   \end{array} \right),
\end{equation}
using the conditions $x=r_{i}$, $y=0$ and where $p_{x}$ is taken \emph{right before} the rebound on the wall. We note that the tangent rebound matrix is a symplectic matrix, which proves that the rebound condition~(\ref{eq:ReboundCondition}) preserves the symplectic two-form, i.e.,
$$
   dx_{0} \wedge dp_{x,0} + dy_{0} \wedge dp_{y,0} = dx \wedge dp_{x} + dy \wedge dp_{y} .
$$

The tangent flow is used to characterize the linear stability of invariant structures like periodic orbits. As for trajectories, their integration is carried out piecewise: the integration is composed of intervals of laser-driven propagation and rebound conditions. Between the rebounds, we integrate the tangent flow given by~\cite{chaosbook}
$$
   d_{t} {\it J} = \nabla {\bf F}\  {\it J},
$$
where~$\nabla{\bf F}$ is the matrix of variations of the flow associated with Hamiltonian~(\ref{eq:RotatedHamilt}). Then, right after a rebound on one wall, the Jacobian matrix is equal to the product of the previous Jacobian matrix right before the rebound, denoted ${\it J}^-$, with the rebound matrix~(\ref{eq:TangentReboundMatrix}): 
Jacobian matrix after the rebound reads 
$$
    {\it J}^{+}={\it J}_R {\it J}^{-}.
$$

\section{Analysis of the dynamics} \label{sec:analysis}

\begin{figure}
  \centering
  \includegraphics[width = .49 \linewidth]{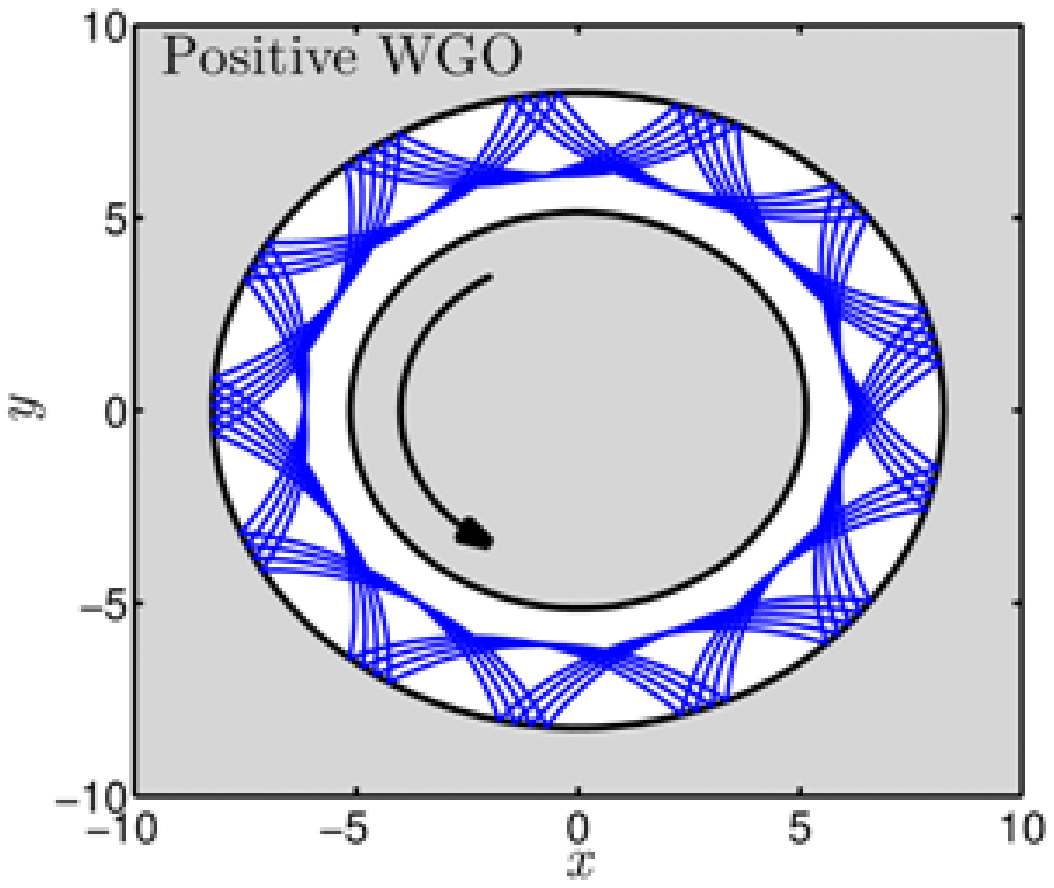}
  \includegraphics[width = .49 \linewidth]{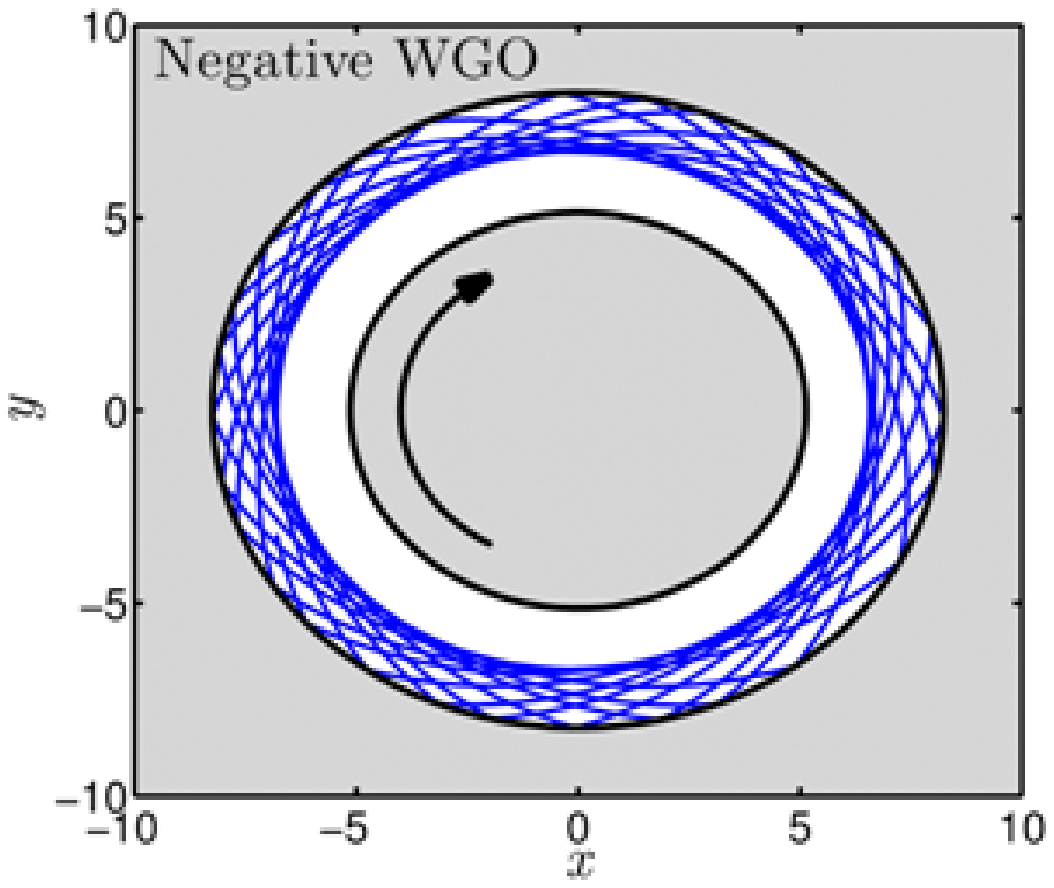}  
  \includegraphics[width = .49 \linewidth]{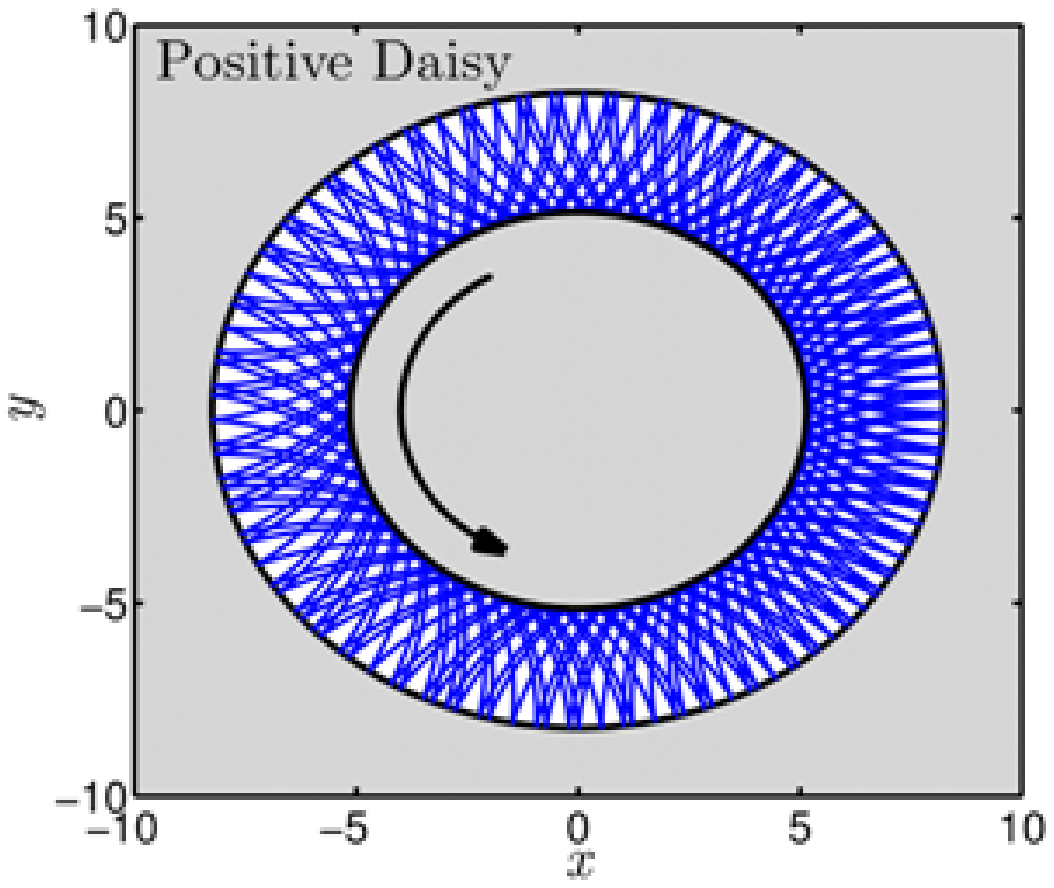}
  \includegraphics[width = .49 \linewidth]{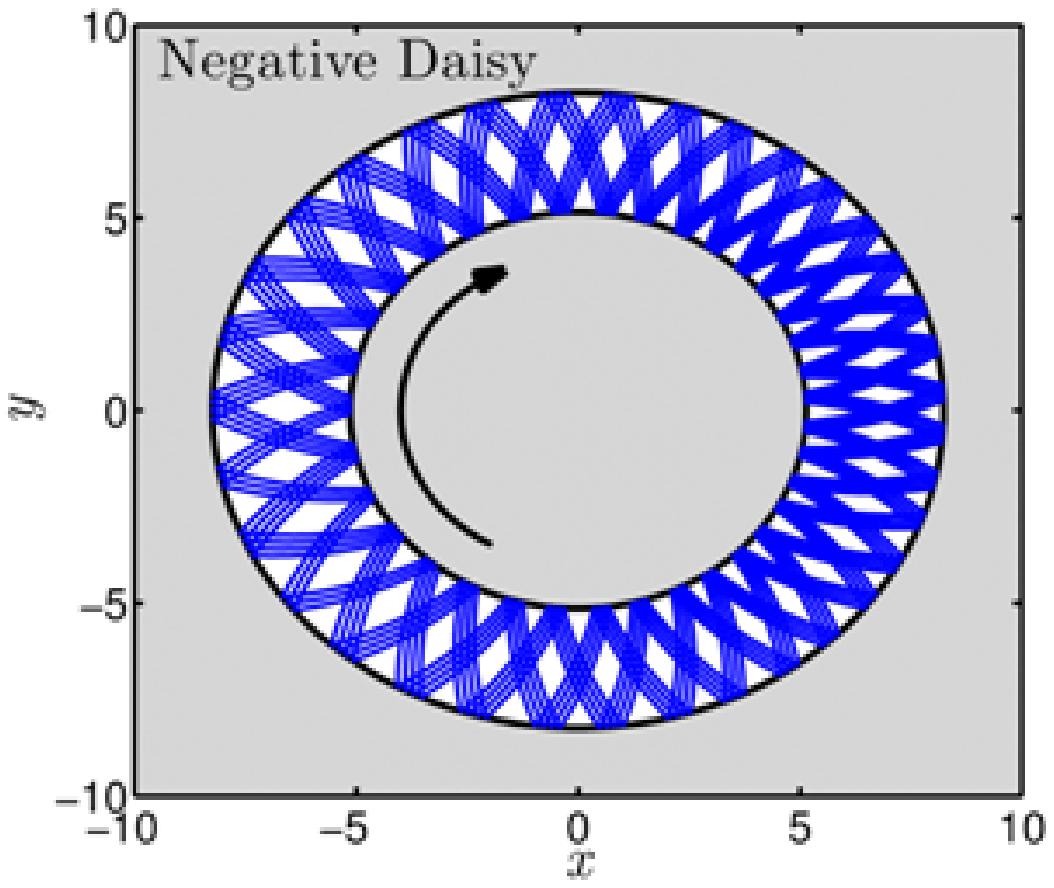}
  \includegraphics[width = .49 \linewidth]{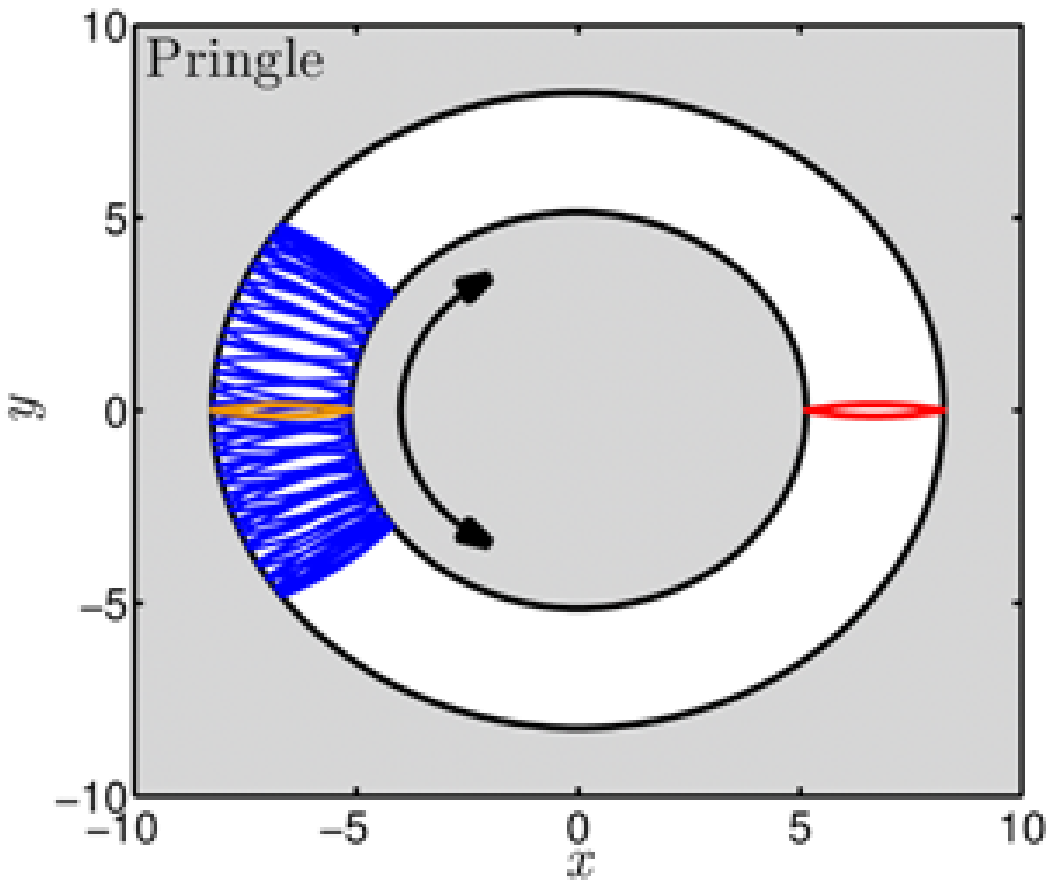}
  \includegraphics[width = .49 \linewidth]{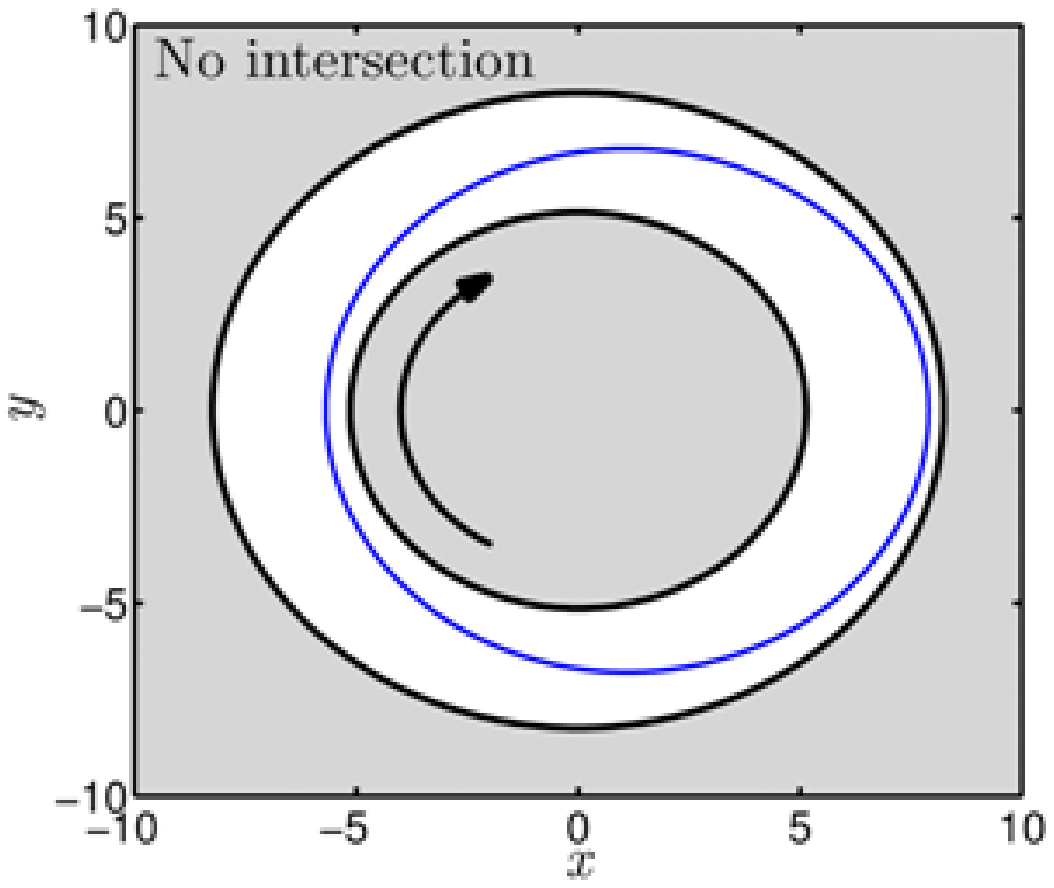}
  \caption{\label{fig:sampleTrajectories}
  Trajectories of the annular billiard for $I=10^{12} \ {\rm W} \cdot {\rm cm}^{-2}$ and $\omega=0.0584$ a.u.  The trajectory type is shown in the top left corner of the panel and $\mathcal{K}=0.35$ (see Fig.~\ref{fig:zvs}) for all panels with the exception of the bottom right.  The arrow shows the direction of travel of the trajectory.  In the bottom left panel the simple two rebound orbits are periodic orbits.  The downfield orbit (orange) is elliptic and the upfield orbit (red) is hyperbolic. The bottom right panel is a trajectory which hits neither wall for $\mathcal{K}=2U_p\approx0.002$ (see Sec.~\ref{sec:noWallHit}).}
\end{figure}

A sampling of typical trajectories of the annular billiard for $I=10^{12} \ {\rm W} \cdot {\rm cm}^{-2}$ is shown in Fig.~\ref{fig:sampleTrajectories}.  These examples illustrate qualitatively the different types of observed trajectories already discussed in Sec.~\ref{sec:intro}.  In the top row we see that the trajectories only hit the outer wall and never the inner wall and are therefore ``whispering gallery orbits'', or WGOs.  In the middle row we show ``daisy orbits''.  They are qualitatively the same, hitting both walls successively and accessing the entire angular distribution of the billiard.  Both WGO and ``daisy orbits'' keep a constant rotational direction, either clockwise or counterclockwise, which we denote negative or positive, respectively.  In the bottom row, left panel we see a ``pringle orbit'', hitting both walls in turn, however, limited only to the downfield region of the billiard.  The simple two rebound trajectories located at both extremes of the upfield and downfiled region of the billiard are periodic orbits.  The leftmost curve is an elliptic periodic orbit (stable) while the rightmost curve is a hyperbolic periodic orbit (unstable).  In the bottom right panel is a trajectory which hits neither wall (see Sec.~\ref{sec:noWallHit}).  In the following sections we connect these trajectories to phase space structures and their stability.

\subsection{Poincar\'{e} Sections} \label{sec:ps}
\begin{figure}
 \centering
 \includegraphics[width = \linewidth]{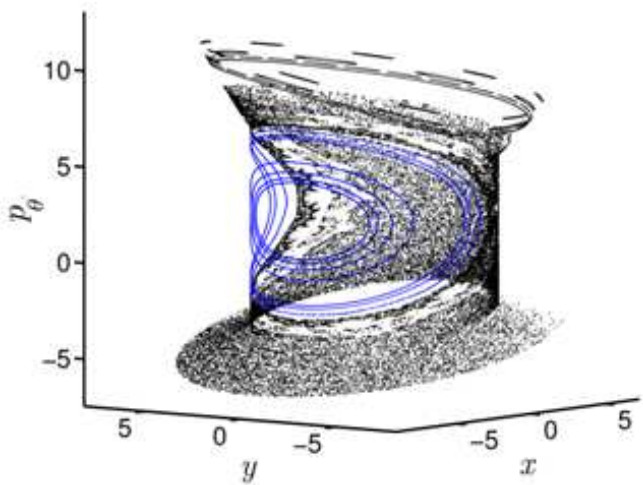}
 \caption{\label{fig:3DPoinc} A Poincar\'{e} section for $I=10^{14} \ {\rm W} \cdot {\rm cm}^{-2}$, $\omega=0.0584$ a.u., and $\mathcal{K}=0.35$.  The blue trajectories are the namesake of ``pringle orbits''.  The corresponding two dimensional projection is displayed in Fig.~\ref{fig:poincSections}, bottom right panel.}
\end{figure}

Since the dynamical system has two degrees of freedom, a convenient way to visualize the dynamical organization of phase space is by Poincar\'{e} sections. Here we consider a Poincar\'e section with equation $p_r=0$ in the direction $\dot{p}_r > 0$.  The rebound condition~(\ref{eq:ReboundCondition}) imposes a discontinuity in the radial momentum $p_r$ at the rebound, such that formally the condition $p_r=0$ is never reached during a rebound.  However, we see the billiard as the limit of an infinitely stiff potential, and a smoother dynamics corresponding to Hamiltonian~(\ref{eq:fullModel_Ham}) would reach $p_r=0$ before changing sign.  As a consequence, we consider the rebounds on the walls as potential candidates for the Poincar\'{e} section.  The electron either rebounds on the outer wall, meaning that the radial momentum changes from positive to negative values or the electron rebounds on the inner wall, so that the radial momentum changes from negative to positive values.  In order to comply with the transverse condition $\dot{p}_r > 0$ we include only the collisions with the inner wall.  Furthermore, because of the rebound condition~(\ref{eq:ReboundCondition})  we note that $p_\theta$, $x=r\cos \theta$, and $y=r\sin\theta$ are continuous under a rebound so that it is equivalent to record their values either directly before or after the rebound.  The Poincar\'{e} section can be represented in several ways.  A three-dimensional plot can be used where we plot $\left(x,y,p_\theta\right)$ or we can make a projection onto the plane $\left(\theta,p_\theta\right)$ using the constant Jacobi constraint. Regardless of which method is chosen there are two types of points on the section: The first kind are points on the inner wall for which $r=r_{\rm in}$ and $p_r$ chosen so as to satisfy the condition on the Jacobi constant ${\cal K}$. The second kind are points for which $p_r=0$ and $r$ is chosen so as to satisfy ${\cal K}$.  In this paper, for the sake of simplicity of our figures, we make use of the projection onto the two dimensional plane $\left(\theta,p_\theta\right)$, however, we show the three dimensional counterpart in Fig.~\ref{fig:3DPoinc} where we have colored ``pringle'' trajectories in blue to illustrate their namesake shape.

\subsubsection{Trajectories which intersect neither wall} \label{sec:noWallHit}
It is natural to ask whether the choice of Poincar\'e section is a good one, i.e.,\ do all trajectories intersect the Poincar\'e section? A small subset of trajectories which are noteworthy for both their peculiarity and the dynamics they showcase, do not intersect the Poincar\'{e} section $p_r=0$. In that spirit we analyze the trajectories which do not collide with either wall. For such trajectories, $x^2+y^2$ needs to remain between $r_{\rm in}^2$ and $r_{\rm out}^2$ at all times. The dynamics of these trajectories is governed by Hamiltonian~(\ref{eq:Ham}).  Using the translation $\tilde{x}=x-E_0/\omega^2$, $\tilde{p_x}=p_x$, $\tilde{y}=y$ and $\tilde{p_y}=p_y-E_0/\omega$, the Hamiltonian is mapped to
$$
\tilde{\cal K}=\frac{p_x^2+p_y^2}{2}-\omega(xp_y-yp_x)+\frac{E_0^2}{2\omega^2},
$$
where we have dropped the tildes for simplicity. The dynamical features no longer depend on the value of $E_0$. The dynamics is better seen in polar coordinates, where the Hamiltonian becomes 
$$
{\cal K}=\frac{p_r^2}{2}+\frac{p_\theta^2}{2r^2}-\omega p_\theta +\frac{E_0^2}{2\omega^2}. 
$$
Since $p_\theta$ is a conserved quantity, the dynamics is that of a particle evolving in a potential equal to $p_\theta^2/(2r^2)$ and the particle will collide with a wall unless $p_\theta=0$. In the case where, $p_\theta=0$, $p_r$ is constant, and it has to vanish so that no collision with the walls takes place.  Therefore the only trajectories which do not hit a wall are circular orbits (since $\dot{r}=p_r=0$). In the original coordinates, these circular periodic orbits are centered around $(x_0,y_0)=(E_0/\omega^2,0)$ and they have a specific Jacobi constant of $2U_p$ where $U_p=E_0^2/(4\omega^2)$ is the ponderomotive energy. For this Jacobi constant, there exist a priori an infinite number of such orbits since the radius is not fixed. The only constraint on the radius is that the circular orbit has to fit inside the annulus. 

Based on the laser parameters $E_0$ and $\omega$, the existence and characterization of such orbits can be divided into several categories. For realistic fullerene parameters, $\left(r_{\rm out}-r_{\rm in}\right)/2<r_{\rm in}$, which is considered here, the analysis can be grouped into four categories.  Different parameters may lead to a different decomposition that can nevertheless be identified in a similar fashion (for instance the third point below may dissappear). 

\begin{enumerate}
  \item If $E_0/\omega^2$ is larger than $r_{\rm out}$, then such trajectories do
not exist because the center is outside the billiard. \\ 
  \item If $E_0/\omega^2$ is in between $r_{\rm in}$ and $r_{\rm out}$, then the
circular orbits are on the right hand side of the inner wall of the annular billiard. \\
  \item If $E_0/\omega^2$ is in between $\left(r_{\rm out}-r_{\rm in}\right)/2$ and $r_{\rm in}$, then such trajectories
do not exist because none of the orbits can be fit inside the allowed region.\\ 
  \item If $E_0/\omega^2$ is smaller than $\left(r_{\rm out}-r_{\rm in}\right)/2$, then the circular orbits
surround the inner wall with a slight shift in the right direction. \\
\end{enumerate} 
The bottom panel of Fig.~\ref{fig:sampleTrajectories} shows a sample trajectory in the fouth category.  In this example, the value $E_0/\omega^2 \approx 1.1$, which is less than ${\rm r_{in}}$ and hence the trajectory surrounds the inner wall, but is shifted slightly to the right.  Except in rare cases (where $E_0/\omega^2$ is equal to $r_{\rm in}$ or $r_{\rm out})$, if such orbits exist, then they exist as a continuous family. A linear stability analysis shows that these orbits are parabolic. In addition, given that $\dot{\theta}=-\omega$ (since $p_\theta=0$), the particle turns clockwise.

Of course, these orbits remain exceptional, in the sense that they only exist at some particular value of the Jacobi constant ${\cal K}=2U_p$. For $\omega=0.0584 \mbox{ a.u.}$, all the orbits hit one wall at least for intensities larger than $I=5.56\times 10^{13}\ {\rm W}\cdot {\rm cm}^{-2}$ or intensities between $2.16\times 10^{13}\ {\rm W}\cdot {\rm cm}^{-2}$ and $1.96\times 10^{12}\ {\rm W}\cdot {\rm cm}^{-2}$. The circular orbits confined in the right hand side of the annulus only exist for intensities between $2.16\times 10^{13}\ {\rm W}\cdot {\rm cm}^{-2}$ and $5.56\times 10^{13}\ {\rm W} \cdot {\rm cm}^{-2}$.  Finally, for intensities lower than $1.96\times 10^{12}\ {\rm W}\cdot {\rm cm}^{-2}$ an infinite family of circular orbits surrounding the inner wall exist. Apart from the examples illustrated in this section all trajectories intersect, an infinite number of times, the Poincar\'{e} section.  Therefore, we can safely keep our definition of the Poincar\'{e} section without missing important dynamics.

\subsubsection{Varying the intensity}
\begin{figure}
  \centering
  \includegraphics[width = .49 \linewidth]{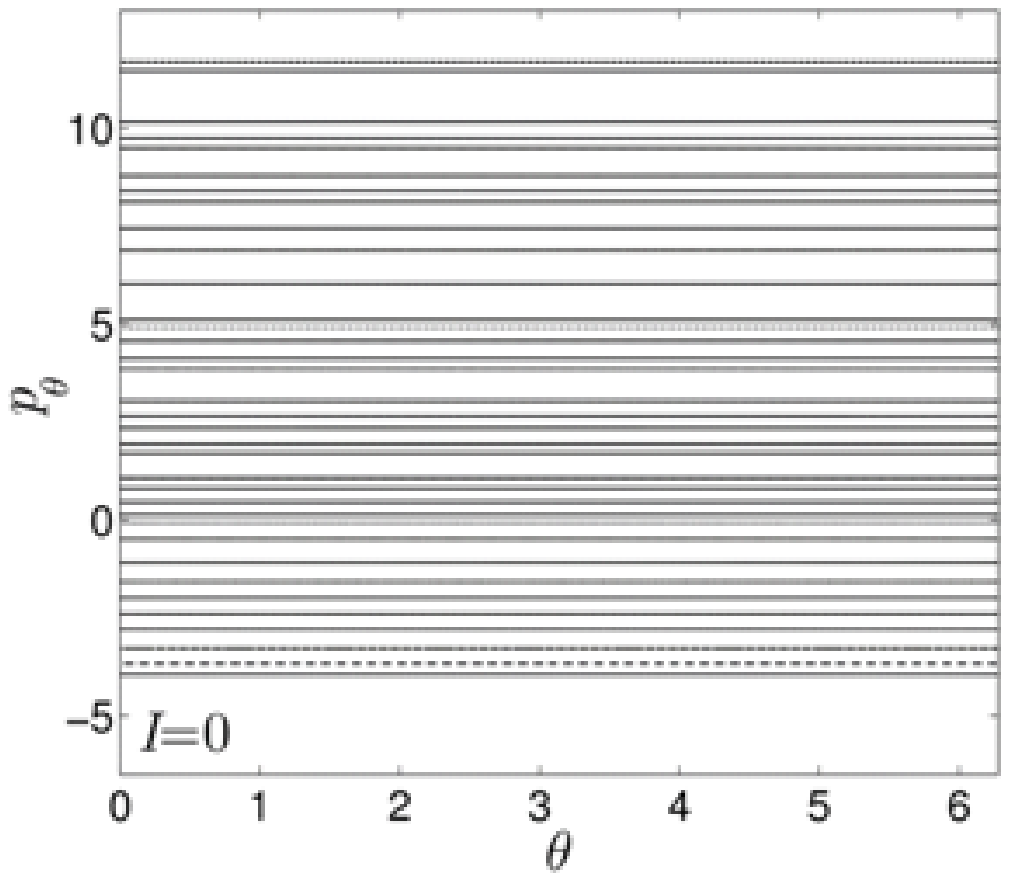}
  \includegraphics[width = .49 \linewidth]{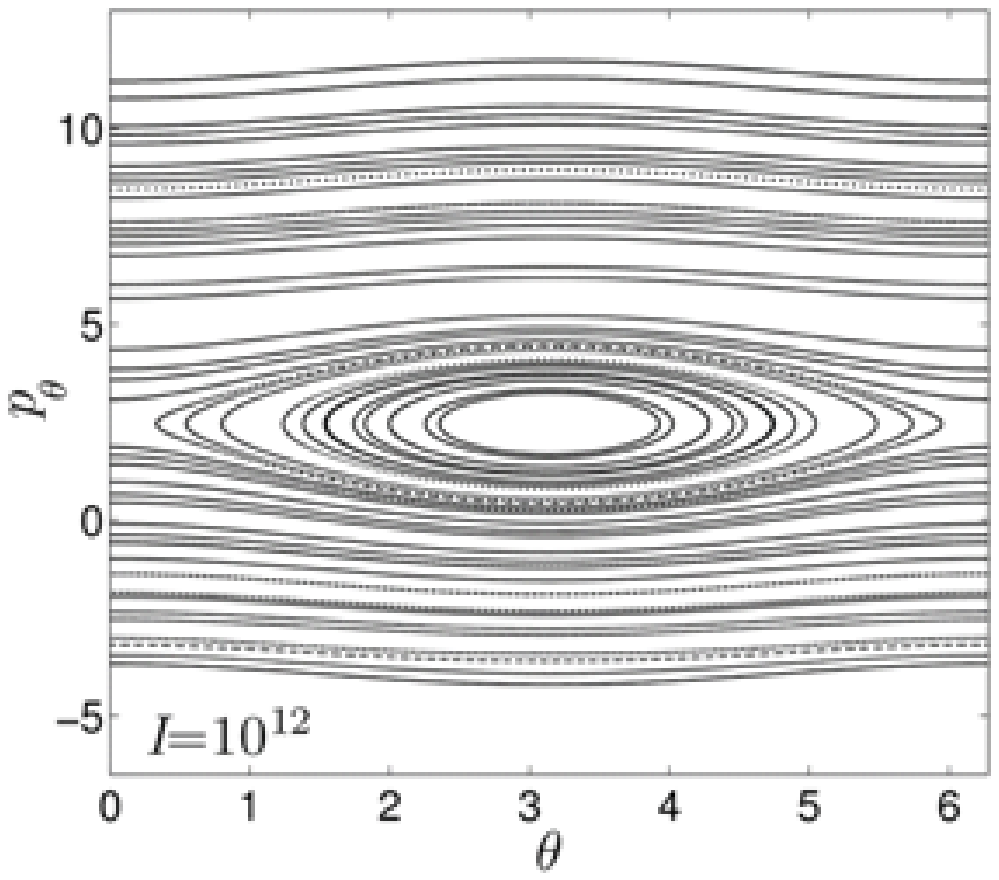}
  \includegraphics[width = .49 \linewidth]{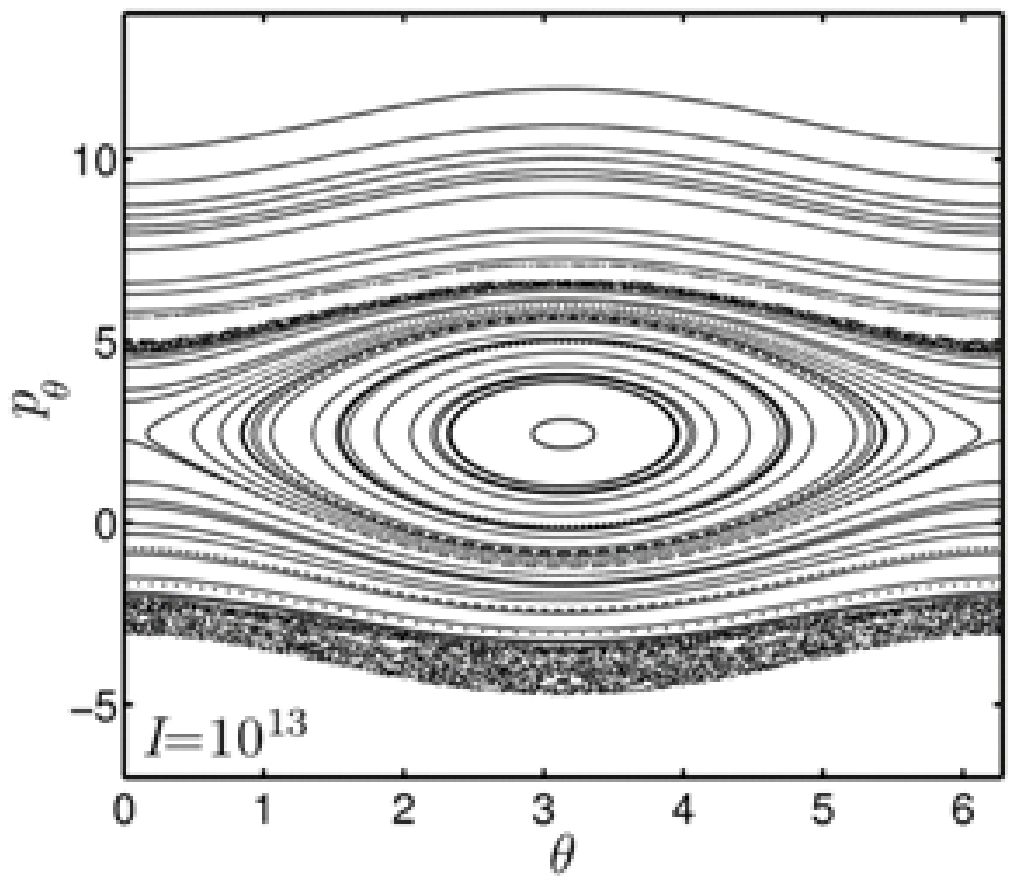} 
  \includegraphics[width = .49 \linewidth]{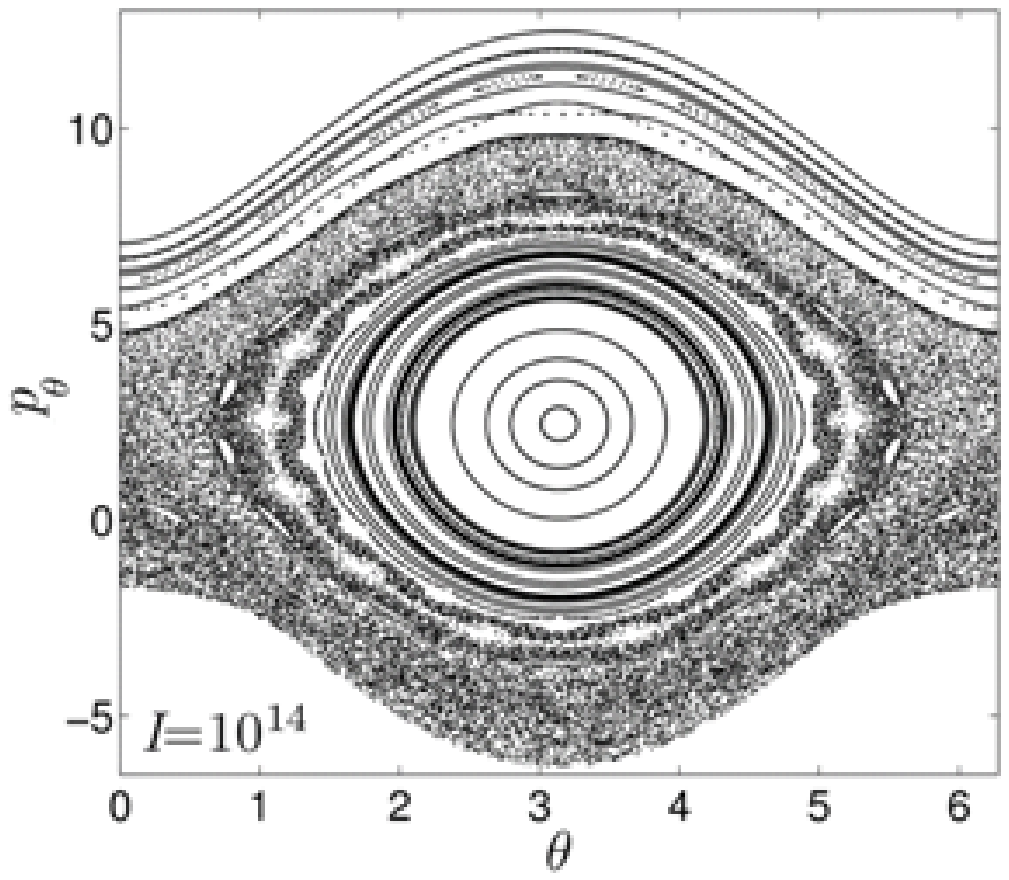}
  \caption{\label{fig:poincSections} 
  Poincar\'e sections for various values of the laser intensity.  Starting in the top left panel we begin with $I=0$ (integrable case) and moving left to right and up to down the intensity is increased.  In all panels $\omega=0.0584$ a.u. and $\mathcal{K}=0.35$.}
\end{figure}

For $E_0=0$, Hamiltonian system~(\ref{eq:Ham}) presents a continuous symmetry by rotation with two degrees of freedom, so it is integrable.  We show the corresponding Poincar\'{e} section in Fig.~\ref{fig:poincSections} (top left panel).  As expected, phase space is foliated by invariant tori.

When $E_0>0$ the system is no longer integrable and some invariant tori are expected to be broken.  According to KAM theory, a large portion of invariant tori persist for $E_0$ small. In Fig.~\ref{fig:poincSections} we show the evolution of phase space as the laser intensity is varied.  With increasing intensity we note the development of a resonance near $p_\theta\approx 2.5$. This resonance corresponds to the aforementioned two-rebound elliptic periodic orbit shown by the left most (downfield) orange curve in the bottom left panel of Fig.~\ref{fig:sampleTrajectories}. This very robust elliptic periodic orbit (situated on the left hand side of the annulus) is extremely important in shaping the overall structure of phase space as intensity increases. Trajectories originating in this region cannot access the entire billiard in a way analogous to the librational motion in a pendulum.  It is this librational motion which yields the ``pringle orbits'' already discussed at the beginning of this section. The behavior in the vicinity of this main resonance can be described roughly in the following way: The resonance is approximately located at $p_\theta=\omega r_0^2$, and this can be seen from the dynamical equation for $\theta$, i.e.,\ $\dot{\theta}=-\omega+p_\theta/r^2$. For a given trajectory, if all the values of $p_\theta$ are larger than $\omega r_{\rm out}^2$, then the trajectory turns counterclockwise. If the values of $p_\theta$ are smaller than $\omega r_{\rm in}^2$ then it turns clockwise. In between, it oscillates between the two tendencies.  

The Poincar\'e sections show that the phase space is highly regular over several decades of laser intensity. Chaotic regions of phase space develop near the hyperbolic periodic orbit which is a rebound between the two walls located in the upfield region, or the right-hand side of the annulus, and is shown in the rightmost curve in the bottom left panel of Fig.~\ref{fig:sampleTrajectories}. Overall the structure of phase space looks very similar to that of a forced pendulum.  In particular, the width of the main resonance zone grows like $\sqrt{E_0}$ or equivalently like $I^{1/4}$. However, we will see that there is a number of discrepancies for which twistless tori are the most significant ones. We readily observe that the lower part of phase space (negative angular momentum) is more chaotic than the upper one (positive angular momentum).  

\subsubsection{Varying the Jacobi value} \label{sec:varyingK}
\begin{figure}
  \centering
  \includegraphics[width = .49 \linewidth]{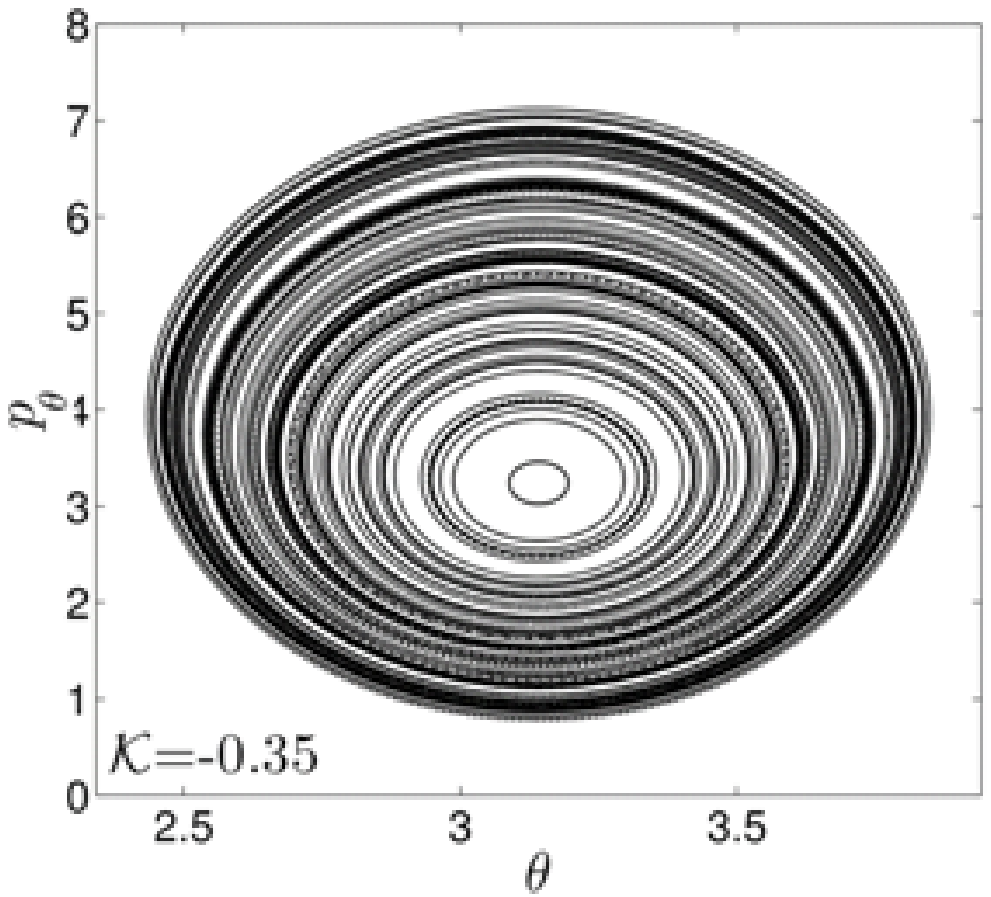}
  \includegraphics[width = .49 \linewidth]{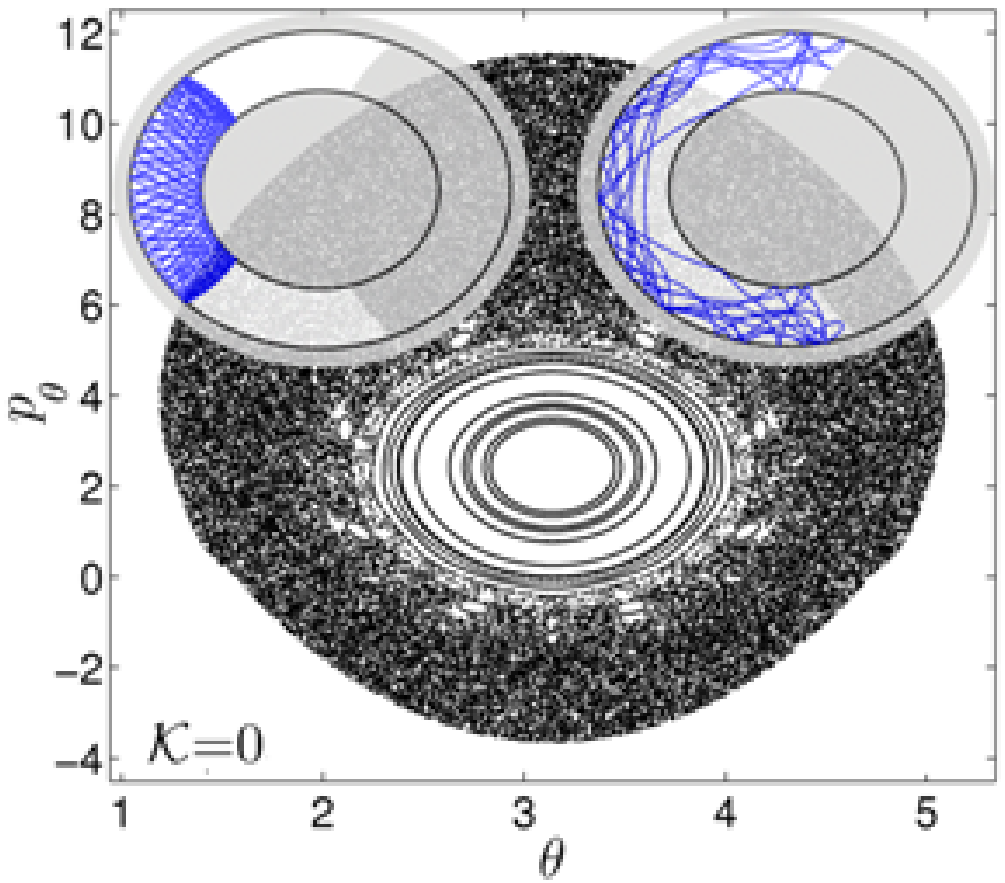}
  \includegraphics[width = .49 \linewidth]{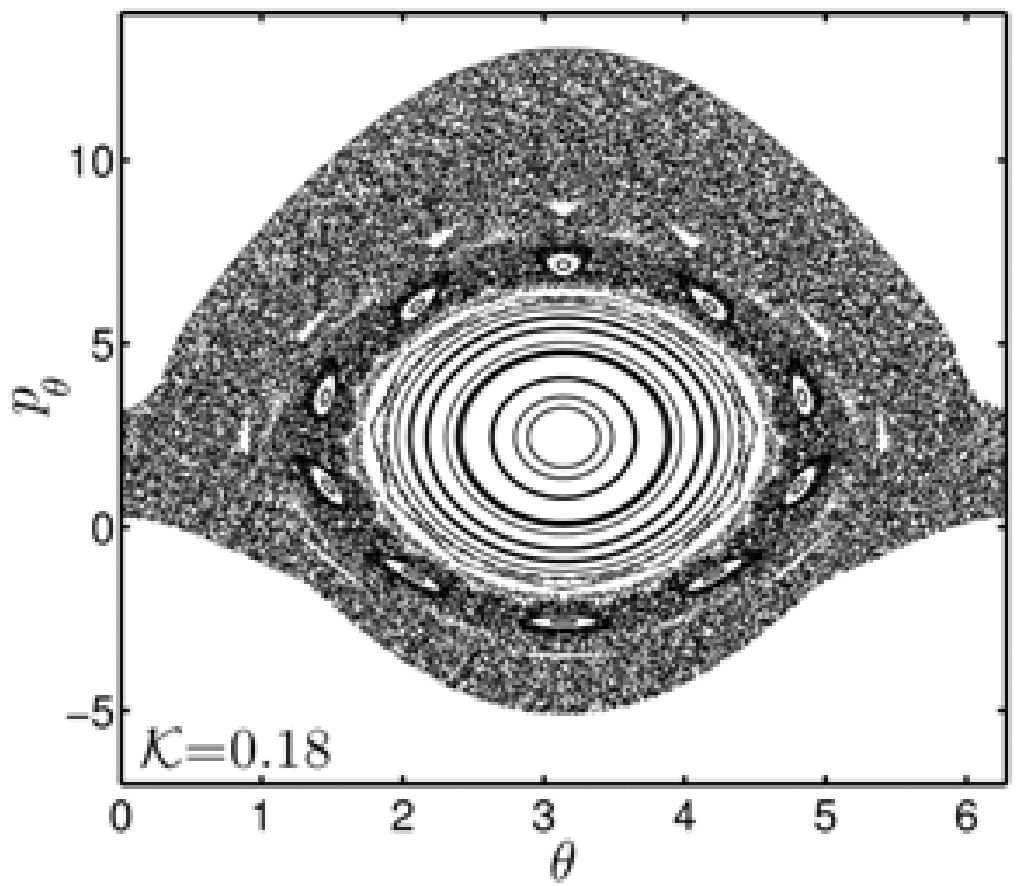} 
    \caption{\label{fig:varyK} 
    Poincar\'e sections for various values of $\mathcal{K}$ with $I=10^{14} \ {\rm W} \cdot {\rm cm}^{-2}$ and $\omega=0.0584$ a.u.  These Jacobi values are the same used in Fig.~\ref{fig:zvs} and Fig.~\ref{fig:accessibleRegions}.  The top right panel includes two trajectories.  The left trajectory is taken from the ``pringle'' region and the right trajectory is taken from the chaotic region.}
\end{figure}

With the intensity fixed at $I=10^{14} \ {\rm W} \cdot {\rm cm}^{-2}$ the Jacobi value can also be varied, keeping in mind that this variation affects the accessible regions of the billiard (see Sec.~\ref{sec:topology}).  We show the corresponding Poincar\'{e} sections in Fig.~\ref{fig:varyK}.  Starting with the top left panel, the Poincar\'{e} section is contained inside the interval $\theta \in \left[2.45, 3.83 \right]$ which agrees with the corresponding panel in Fig.~\ref{fig:accessibleRegions}, note also that the dynamics is highly regular. Besides, since the inner wall is not accessible for this Jacobi value, points on the section result directly from the condition that $p_r=0$: A typical trajectory hits the outer wall but never reaches the inner one. For the top right panel, the inner wall is now accessible (see top right panel of Fig.~\ref{fig:accessibleRegions}) and the dynamics shows a mixed chaotic and regular behavior. In the regular region of the Poincar\'{e} section the trajectories are regular ``pringle orbits'', whereby the turning points in $\dot{\theta}$ are due to the dynamics (and not to the geometry of the billiard).  Recall that the regular region does not span the entire region of accessible $\theta$ values.  However, the trajectories originating in the chaotic region experience turning points in $\dot{\theta}$ because they hit the artificial walls imposed by the choice of Jacobi value.  In the bottom panel the entire inner wall is accessible and only a small portion of the outer wall is inaccessible. The dynamics is still mixed, composed of a regular region with ``pringle orbits'', and a highly chaotic region where the trajectories hit the virtual walls imposed by the geometry of the configuration space.  WGOs are not possible for these values of Jacobi constant. In fact, WGOs appear only when the entire annulus is accessible, e.g. in the bottom right panel of Fig.~\ref{fig:poincSections}.

\subsection{Partitioning of phase space}
Because of the dimensionality of the billiard (two degrees of freedom) invariant KAM tori constitute barriers of transport which confine the (chaotic) dynamics to distinct regions of phase space. However, they are not robust enough to partition phase space at sufficiently high value of the intensity. Here, our analysis reveals the existence of much more robust invariant objects, namely twistless tori which are particularly relevant for the organization of phase space~\cite{Morr09,Negr96}, since they partition phase space at relatively high intensities into regions where the different types of trajectories occur. We next introduce a diagnostic tool for finding these tori.

\subsubsection{Frequency Analysis} \label{sec:freqAnalysis}
Frequency analysis~\cite{Lask93} is a practical tool in Hamiltonian systems for analysis of the dynamics. For integrable systems written in action-angle variables $({\bf A},\bm{\varphi})$, the method consists of plotting the frequency ${\bm \omega}({\bf A})=\partial H_0/\partial {\bf A}$ as a function of ${\bf A}$, which is expected to be smooth for (sufficiently smooth) integrable systems. For nearly integrable systems, the frequency is computed by a windowed Fourier transform of a chosen observable. It is computed for an ensemble of trajectories, and plotted, for instance, as a function of the initial value of the action. From this analysis it is possible to identify elliptic and hyperbolic islands, regular regions filled by KAM tori, and chaotic regions by their respective unique signatures. The elliptic islands are expressed as constant frequency plateaus, the hyperbolic orbits by cusps in the frequency, regular regions as apparently continuous curves, and chaotic regions as non-smooth sections. Frequency analysis can also identify regions where the twist condition is not satisfied, i.e., when ${\bm \omega}$ is no longer a monotonous function of the action. In this case, the standard twist condition for the standard KAM theorem is not satisfied, and it gives rise to a new taxonomy of dynamical mechanisms, like separatrix reconnection and twistless tori~\cite{Morr09,Negr96}. 

\begin{figure}
 \centering
 \includegraphics[width = \linewidth]{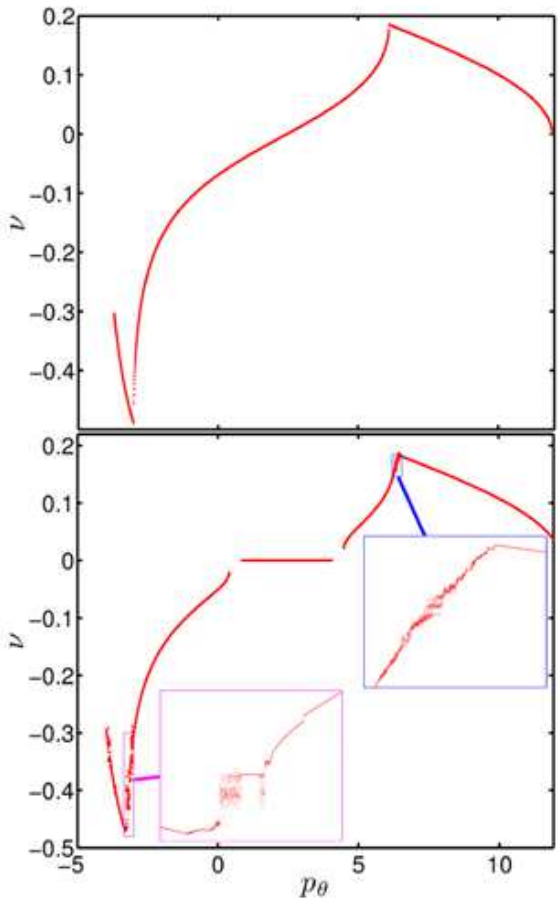}
 \caption{\label{fig:freqMap}  
 Frequency analysis in the integrable case ($I=0$, upper panel) and non-integrable case ($I=10^{12} \ {\rm W} \cdot {\rm cm}^{-2}$, lower panel).  The insets display the frequency map around the extrema.  In both panels $\mathcal{K}=0.35$ and $\omega=0.0584$ a.u.}   
\end{figure}

In Figure~\ref{fig:freqMap} we plot the frequency as a function of the initial momentum $p_\theta$. The ensemble of trajectories are the series of points on the Poincar\'e section with initial conditions $\theta=\pi$ and various $p_\theta$ while imposing the Jacobi constraint to compute the other variables. The motivation is that for the integrable case $p_\theta$ is a conserved quantity. The frequency analysis of the integrable case (Fig.~\ref{fig:freqMap}, upper panel) shows that the frequency is a continuously varying function of $p_\theta$, as is expected.  The most interesting feature is that the frequency map does not change monotonically with $p_\theta$.  This implies the existence of twistless tori located at the extrema of the frequency map, at $p_\theta \approx -3.03$ and $p_\theta \approx 6.11$.  In the non-integrable case (Fig.~\ref{fig:freqMap}, lower panel), we see the appearence of a plateau which corresponds to the ``pringle orbits''.  In addition we notice some chaotic features in the insets, even if the overall behavior seems to be quite regular at this intensity.  Furthermore, the frequency map still exhibits two extrema, again which correspond to twistless tori. A closer look at the frequency map around $p_\theta=-3.2$ and $p_\theta=6.5$ (insets) reveals a rich dynamics with a succession of elliptic and hyperbolic orbits, even if the overall behavior seems to be quite regular.

\subsubsection{Twistless tori}
\begin{figure}
 \centering
 \includegraphics[width = \linewidth]{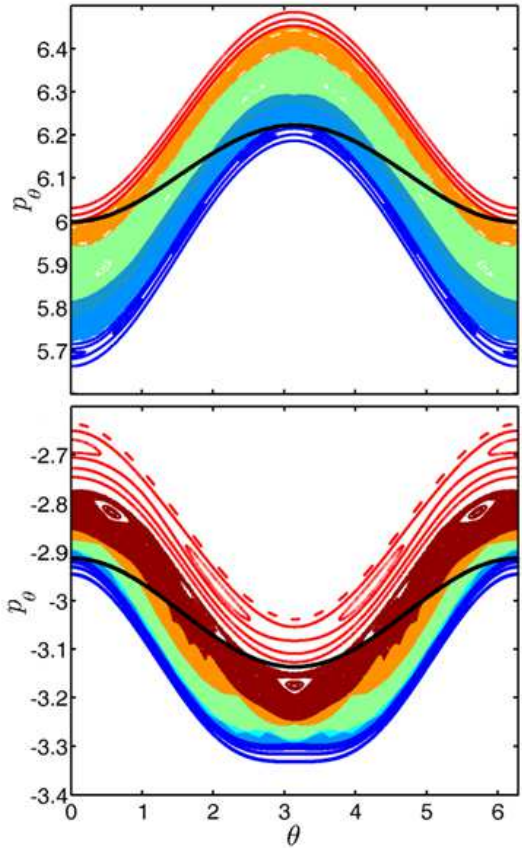}
 \caption{\label{fig:meanderingTori} 
 Poincar\'{e} sections of parts of phase space near the minimum (lower panel) and maximum (upper panel) of Fig.~\ref{fig:freqMap}.  In both panels, filled in layers correspond to each chaotic region separated by a twistless torus.  The parameters are chosen the same as in Fig.~\ref{fig:freqMap} ($I=10^{12} \ {\rm W} \cdot {\rm cm}^{-2}$, $\mathcal{K}=0.35$, and $\omega=0.0584$ a.u.)}    
\end{figure}

\begin{figure}
 \centering
 \includegraphics[width = \linewidth]{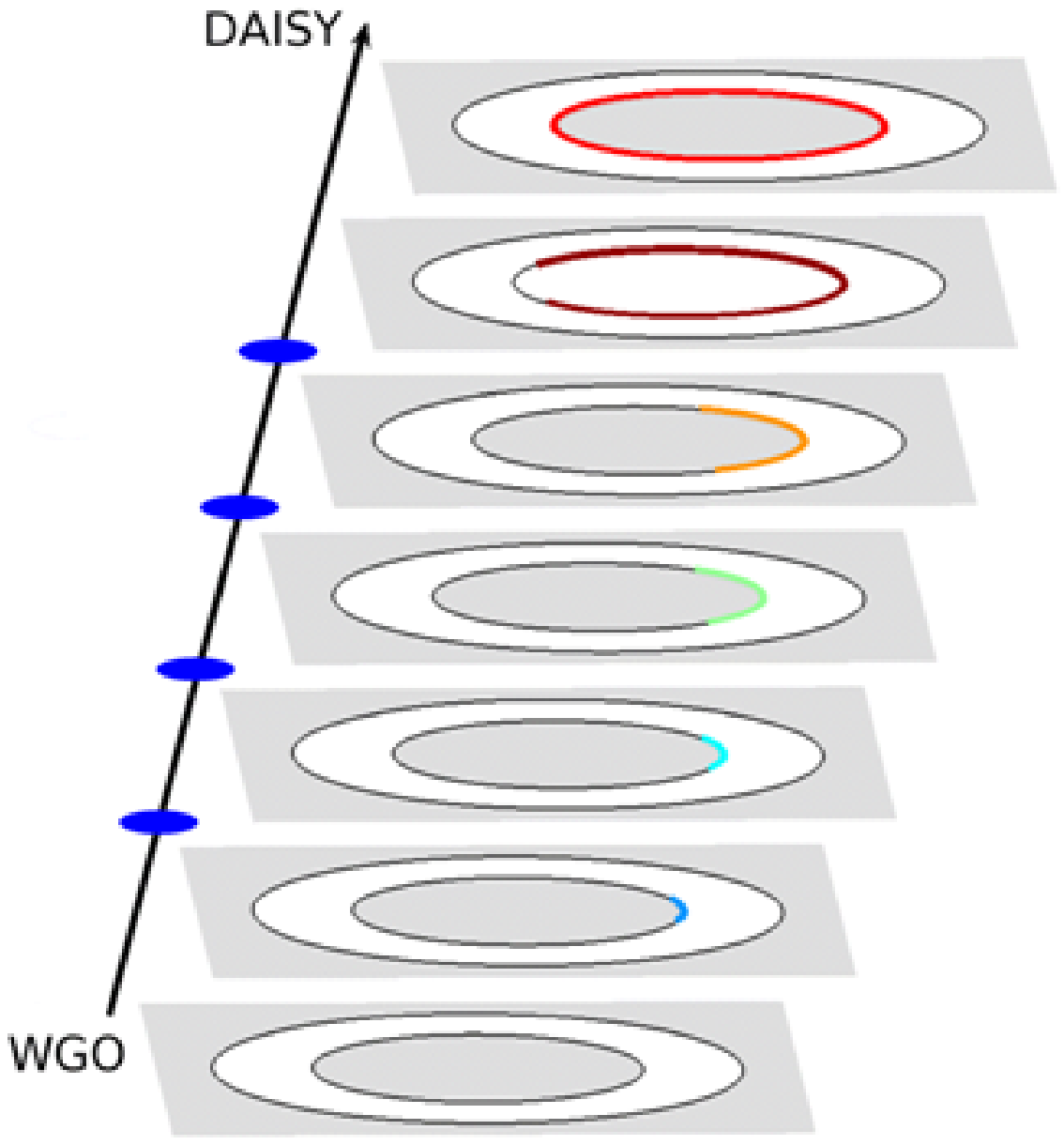}
 \caption{\label{fig:transition} 
 The transition scenario from WGOs to ``daisy orbits'' is schematically represented via a graphic showing successive regions of the inner wall which can be visited by the electron.  The color of the accessible region of the inner wall corresponds to the color of the chaotic regions in the bottom panel of Fig.~\ref{fig:meanderingTori}. The blue ovals represent the twistless tori which separate the different chaotic regions.}    
\end{figure}

The aforementioned twistless tori are visualized by high resolution Poincar\'{e} sections. Figure~\ref{fig:meanderingTori} shows Poincar\'{e} sections with initial conditions nearby $p_\theta \approx 6.5$ (upper panel) and $p_\theta \approx -3.2$ (lower panel), corresponding to the local maxima in Fig.~\ref{fig:freqMap}.  Both Poincar\'{e} sections are similar in that they show well-developed chaotic regions sandwiched between two regular regions.  The regular regions correspond to WGOs, shown in blue, and ``daisy orbits'', shown in red.  Likewise, the chaotic region in both panels exhibits a stratification whereby a trajectory originating in one of these regions remains there and cannot pass to another chaotic region. The stratification is not due to KAM tori since KAM tori come in families. Instead it is caused by the existence of twistless tori, which, having dimension two, can partition phase space. In particular in the lower panel of Fig.~\ref{fig:meanderingTori}, we recognize one of the signatures of twistless tori which is the meandering behavior (see for instance the interface between the orange and the dark red chaotic regions). The black line, superimposed over both panels, separates the two different possible ways of intersecting the Poincar\'{e} section.  Points on the section above (resp. below) the black line for the upper (lower) panel are standard intersections of the flow with the Poincar\'{e} section in the sense that $p_r$ changes sign smoothly before and after the intersection. Points below (resp. above) the black curve for the upper (lower) panel are collisions with the inner wall where the sign of $p_r$ changes due to the rebound condition (see Sec.~\ref{sec:ps}).  As expected, all the WGO trajectories are below the black curve in the lower panel (since none of their points intersect the inner wall). The entire WGO region is regular. We also notice that all the ``daisy orbits'' are above the black curve, and this region is also mostly regular. Between these two regions is a strongly chaotic region with very few elliptic islands. Each of the chaotic regions (in both panels) has a portion above and below the black line.  The range of $\theta$ where each chaotic region exists above (below) the black line gives the accessible region of the inner wall to the electron. Figure~\ref{fig:transition} illustrates how the accessible region of the inner wall changes as one moves from WGOs to ``daisies'' in the lower panel of Fig.~\ref{fig:meanderingTori}.  Beginning with the WGOs, shown in blue, which have no points above the black line (no accessible region on the inner wall) we increase $p_\theta$ to the light blue region there is a small range of $\theta$ values for which the chaotic region is above the black line.  The $\theta$ values are centered about $\theta=0$ and they correspond to the area along the inner wall which is accessible to the electron. Moving again upwards in $p_\theta$ we pass by several more chaotic regions, each having a wider range of $\theta$ above the black line which allows for a wider range of the inner wall to be visited by the electron.  The inner wall becomes more accessible until finally reaching the ``daisy'' region, shown in red. At this point, all intersections of the Poincar\'{e} section are above the black line covering $\theta \in \left[0,2\pi\right]$ and therefore the entire inner wall is accessible. The four twistless tori which exist in this chaotic region are responsible for the discrete transition from WGOs to ``daisy orbits''.  Likewise, a simiar feature can be observed for the upper panel of Fig.~\ref{fig:meanderingTori}, nearby $p_\theta = 6.5$, however there are fewer chaotic regions and hence fewer twistless tori. 

\section*{Conclusions and outlook}
The motivation of our work stems from recent photoionization experiments in strong field of atoms and molecules in both circular and linear polarized light where a significant variation of the yields with polarization was observed~\cite{Hert09}.  We propose a rather simple dynamical model for the motion of a valence electron inside the valence shell of fullerene ${\rm C}_{60}$, namely an annular billiard. We have investigated the dynamics when this electron is subjected to a circularly polarized laser field. We have shown that it exhibits three distinct types of trajectories: ``whispering gallery'', ``daisy'' and ``pringle'' orbits.  These trajectories are found in distinct, identifiable regions of phase space for a wide range of laser intensity and Jacobi values.  They are kept characteristically segregated from each other by the existence of twistless tori which partition phase space.  The twistless tori are identified through a frequency analysis and are confirmed by generation of high resolution Poincar\'{e} sections. These twistless tori, blessed with high stability, exist in chaotic regions where KAM tori have been broken by the strong laser field.  Because of the barriers they create, twistless tori, allow for a transition scenario from WGO orbits to ``daisy orbits'' in both rotational directions, positive and negative.  

\acknowledgments
The authors would like to acknowledge P.~J. Morrison and L.~A. Bunimovich for enlightening discussion.  C.C. and F.M. acknowledge financial support from the CNRS. F.M. acknowledges financial support from the Fulbright program. This work is partially funded by NSF.


\end{document}